\documentclass[twocolumn,reprint,amsmath,amssymb,aps,prl]{revtex4-2}

\usepackage{graphicx}
\usepackage{braket}
\usepackage{xcolor}
\usepackage{hyperref}
\usepackage{comment}
\usepackage{placeins}

\newcommand{\blue}[1]{\textcolor{black}{#1}}
\newcommand{\black}[1]{\textcolor{black}{#1}}

\newcommand{\uvec}[1]{\hat{\mathbf{#1}}}

\newcommand{\phat}{\uvec{\boldsymbol{\ell}}}                
\newcommand{\nhat}{\uvec{\Omega}}                     
\newcommand{\qhat}{\uvec{\boldsymbol{m}}}

\newcommand{\tlh}{\tilde{h}}

\newcommand{\I}{\mathcal{I}}
\newcommand{\J}{\mathcal{J}}

\usepackage{lineno}

\begin{document}

\title{Polarized Anisotropic Stochastic Gravitational Wave Background Search with Ground-Based Detector Networks}

\author{Töre Boybeyi}
\email{boybe001@umn.edu}
\affiliation{School of Physics and Astronomy, University of Minnesota, Minneapolis, MN 55455, USA}
\author{Vuk Mandic}
\email{vuk@umn.edu}
\affiliation{School of Physics and Astronomy, University of Minnesota, Minneapolis, MN 55455, USA}

\date{\today}

\begin{abstract}
Gravitational waves admit a Stokes decomposition into intensity ($I$), circular polarization ($V$), and linear polarization ($Q$, $U$), analogous to Cosmic Microwave Background (CMB) polarimetry. We implement a full-Stokes maximum-likelihood SGWB map-making analysis for ground-based detector networks, promoting the standard cross-correlation data products used in existing pipelines to a joint reconstruction of $I$, $V$, $Q$, $U$. Applied to LVK O3 data, \black{we place the first constraints} on the polarized angular spectra $C^{VV}_\ell$, $C^{EE}_\ell$,
$C^{BB}_\ell$ and $|C^{IV}_\ell|$. We show that an intensity-only model is biased when polarized sky components are present, since the detector-network Fisher inner product does not generally make the Stokes responses orthogonal. For transient CBC foregrounds, polarized shot noise is not parametrically suppressed relative to ordinary CBC intensity shot noise. The full Stokes framework separates the Stokes sectors while providing access to polarized anisotropies invisible to conventional intensity-only searches.
\end{abstract}

\maketitle

\textit{Introduction.} The direct observation of gravitational waves (GWs) has opened a new window on the Universe, with hundreds of compact binary coalescence (CBC) candidates detected by the LIGO/Virgo/KAGRA (LVK) collaboration~\cite{LIGOScientific:2026wfs,LIGOScientific:2025slb,KAGRA:2021vkt,LIGOScientific:2016aoc,LIGOScientific:2017vwq} and evidence for a stochastic gravitational wave background (SGWB) at nanohertz frequencies emerging from pulsar timing arrays~\cite{NANOGrav:2023gor,NANOGrav:2020bcs,EPTA:2023fyk,reardon2023search,Xu:2023wog}. As sensitivity improves, characterizing the SGWB beyond its spectrum to include angular anisotropies and polarization becomes both feasible and scientifically compelling~\cite{KAGRA:2021kbb,LIGOScientificCollaborationtheVirgoCollaboration:2025woc,LIGOScientific:2026mjf,Cusin:2017fwz,Jenkins:2018nty,LIGOScientific:2018czr,Callister:2017ocg,Seto:2008sr,Kato:2015bye,Allen:1996gp,Allen:1997ad,Romano:2016dpx,NANOGrav:2021ini,NANOGrav:2023tcn,Mentasti:2023gmg}. Because the SGWB is sourced by both astrophysical populations and possible primordial relics, its spectrum, anisotropy, and polarization encode information about compact binary demographics and large scale structure, as well as early Universe dynamics such as phase transitions or inflation~\cite{LIGOScientific:2025kry,NANOGrav:2023hvm,LIGOScientific:2025pvj,NANOGrav:2023hfp,Maggiore:1999vm,Caprini:2018mtu,Caldwell:2022qsj,LIGOScientific:2026ctl}.

Just as the cosmic microwave background (CMB) is fully characterized by its Stokes parameters $\{I, Q, U, V\}$, the SGWB possesses an analogous polarization structure encoding qualitatively new information~\cite{Seto:2008sr,Callister:2017ocg,Kato:2015bye,Gair:2014rwa,Gair:2015hra}. The intensity anisotropy traces the spatial distribution of GW sources and large scale structure~\cite{Cusin:2017fwz,Jenkins:2018nty}, circular polarization ($V$) probes parity violation in the gravitational sector~\cite{Seto:2008sr,Callister:2017ocg,Smith:2016jqs}, and linear polarization ($Q$, $U$) constrains the orbital geometry of the source population. Existing anisotropic ground-based SGWB map-making searches have reconstructed only the intensity channel~\cite{Ballmer:2005uw,Mitra:2007mc,Thrane_2009,Romano:2015uma,KAGRA:2021mth,LIGOScientific:2025kbu}, leaving the polarized Stokes fields unmodeled. \black{Unlike the vector and scalar polarizations, which require physics beyond general relativity, the tensor polarization is produced by any finite population of sources.}

In this \textit{Letter}, we introduce a full Stokes SGWB search framework for ground-based networks that jointly reconstructs $I,V,Q,U$ from the standard cross-correlation data. \black{The polarized sky therefore becomes an observable with the data that existing searches already produce, without a new measurement}. We apply it to LVK O3 data and establish \black{the first constraints} on the anisotropic polarization of the SGWB. We validate the framework through simulations and demonstrate how unmodeled polarization can bias intensity-only anisotropy estimates.

\textit{Formalism.} We work in the frequency domain and expand the GW strain arriving from sky direction $\nhat$ in terms of the tensor polarization basis $e^{(s)}_{ij}(\nhat)$ (see Supplemental Material (SM)\blue{~\cite{SM}} S-I):
\begin{equation}
    \tlh_{ij}(f,\nhat) = \sum_{s \in \{\pm2\}} e^{(s)}_{ij}(\nhat)\, \tlh^{(s)}(f,\nhat),
\end{equation}
where the sum is restricted to the two tensor helicities $s=\pm2$; vector and scalar modes, which have been targeted in separate stochastic-polarization searches~\cite{LIGOScientific:2018czr,LIGOScientificCollaborationtheVirgoCollaboration:2025woc}, are set to zero throughout. We expand the two-point strain correlation function as:
\begin{equation}
\begin{split}
    &\big\langle \tilde{h}^{(s_1)}(f_1,\nhat_1)\,\tilde{h}^{*(s_2)}(f_2,\nhat_2)\big\rangle = \\
    &\quad \frac{1}{2}\,\delta(f_1-f_2)\,\delta^{(2)}(\nhat_1-\nhat_2)\mathcal{H}_{s_1 s_2}\!\left(f_1;\nhat_1\right), 
    \label{eq:GW_strain_correlations_helicity}
\end{split}
\end{equation}
where $\mathcal{H}$ contains the standard gravitational Stokes parameters~\cite{Romano:2016dpx}:
\begin{equation}
    \mathcal{H}_{s_1 s_2}(f,\nhat) = \frac{1}{2}
    \begin{pmatrix}
    I+V & P^- \\[4pt]
    P^+ & I-V
    \end{pmatrix}\!(f,\nhat). 
    \label{eq:H_helicity_matrix}
\end{equation}
and we assumed different directions to be uncorrelated. The rows are labeled by $s_1 \in \{+2,-2\}$ and the columns by $s_2 \in \{+2,-2\}$. The linear polarization fields $P^\pm \equiv Q\pm iU$ transform as $P^\pm \to e^{\mp 4 i\psi}P^\pm$ under a local basis rotation by $\psi$, identifying them as spin-$\pm4$ fields (spin-4 arising from the product of two spin-2 tensors), whereas $I$ and $V$ are spin-$0$. When expanded in spin-weighted spherical harmonics ${}_{\pm4}Y_{\ell m}$, the linear polarization coefficients $a^\pm_{\ell m}$ separate into parity-even (gradient, $E$) and parity-odd (curl, $B$) components, $a^E_{\ell m} = -(a^+_{\ell m}+a^-_{\ell m})/2$ and $a^B_{\ell m} = i(a^+_{\ell m}-a^-_{\ell m})/2$, in direct analogy with CMB polarimetry~\cite{Gair:2015hra,AnilKumar:2023hza}. Since spin-$\pm4$ harmonics exist only for $\ell\geq4$, the $E$ and $B$ spectra also begin at $\ell=4$.

We consider a detector $\I$ at position $\vec x_{\I}$, whose output $\tilde{s}_{\I}(f;t)$ during a time-segment labeled by $t$ combines instrumental noise $\tilde{n}_\I(f;t)$ with the sky integrated projection of the GW strain onto the detector geometry~\cite{Romano:2016dpx}. For each baseline $\I\J$ and Stokes parameter $X\in\{I,V,P^\pm\}$, one defines direction dependent overlap reduction functions (ORFs) $\Gamma^X_{\I\J}(f,\nhat;t)$ built from the antenna patterns $F^{+,\times}_\I$ and the baseline phase factor~\cite{Chu:2020qiw,Liu:2022umx,AnilKumar:2023yfw} (SM \ref{sec:conventions}).

We compute the one sided cross spectrum $\hat C_{\I\J}(f;t)\equiv 
\tfrac{1}{\tau}\tilde s_\I(f;t)\tilde s^*_\J(f;t)$ from short 
Fourier segments of duration $\tau$. We define the $\Omega$-normalized Stokes fields
$\Omega_X(f,\nhat) \equiv \frac{2\pi^2 f^3}{3H_0^2}\,X(f,\nhat)$ for 
$X \in \{I,V,Q,U\}$. Assuming a power-law spectrum $\Omega_X(f,\nhat) = H(f)\,\Omega_X(\nhat)$ with $H(f) = (f/f_{\rm ref})^\alpha$, we expand the angular dependence in a spatial basis indexed by $\kappa$ ($\kappa = \ell m$ in the spherical harmonic (SH) basis or a pixel index $p$ in a pixel basis):
\begin{equation}\label{eq:stokes_harm}
    \Omega_X(\nhat) = \sum_\kappa \Omega^X_\kappa\, 
    \Psi^X_\kappa(\nhat)
\end{equation}
where $\Psi^X_{\ell m}= {}_{s_X}Y_{\ell m}(\nhat)$ in the SH basis ($s_X=0$ for $X=I,V$ and $s_X=\pm 4$ for $P^{\pm}$), while in the pixel basis $\Psi^X_{p}$ equals one in pixel $p$ and zero elsewhere. Following the standard radiometer and maximum-likelihood SGWB map-making formalism~\cite{Ballmer:2005uw,Mitra:2007mc,Thrane_2009,Suresh:2020khz}, the estimator for the (clean map) anisotropy coefficients is $\hat{\boldsymbol{\Omega}} = \mathcal{F}^{-1}\hat{\mathcal{X}}$, with
\begin{align}
    \mathcal{F}^{XY}_{\kappa \kappa'} &= \sum_{\substack{
    \I<\J \\ f,\, t}} \frac{[\tilde\Gamma^{X}_{\I\J;\kappa}]^{*}\,
    \tilde\Gamma^{Y}_{\I\J;\kappa'}}
    {\bar P_\I(f,t)\,\bar P_\J(f,t)}, 
    \label{eq:fish1}\\
    \hat{\mathcal{X}}^X_{\kappa} &= \sum_{\substack{\I<\J \\ f,\, t}} 
    \frac{[\tilde\Gamma^{X}_{\I\J;\kappa}]^{*}\, 
    \hat C_{\I\J}(f;t)}
    {\bar P_\I(f,t)\,\bar P_\J(f,t)}, 
    \label{eq:fish2}
\end{align}
where $\bar P_\I(f,t)$ are the estimated noise power spectral densities (PSDs), \black{with the windowing, coarse graining and normalization factors of the segmented Fourier analysis absorbed in their definition~\cite{KAGRA:2021kbb}},
$\tilde\Gamma^{X}_{\I\J;\kappa}(f,t) \equiv \frac{3H_0^2}{4\pi^2
f^3} H(f)\int
d\nhat\,\Gamma^X_{\I\J} (f,\nhat;t)\,\Psi^X_\kappa(\nhat)$ are the
$\Omega$-normalized responses projected onto the spatial basis, and the sums run over all baselines, frequency bins, and time segments.
The Fisher matrices $\mathcal{F}$ can be ill-conditioned and, when needed, we use a regularized pseudoinverse with poorly measured eigenmodes removed.

A key consequence of the joint Stokes treatment is the identification of a systematic bias in the standard intensity-only analysis. Because the dirty map
$\hat{\mathcal{X}}^I$ receives contributions from all Stokes parameters, 
the $I$-only clean map $\hat\Omega^I_{\kappa} = \sum_{\kappa'}
(\mathcal{F}^{II})^{-1}_{\kappa\kappa'}\,\hat{\mathcal{X}}^I_{\kappa'}$ 
cannot distinguish polarization from intensity. The Stokes fields are physically distinct, but their detector responses are not generally orthogonal under the network Fisher inner product. Its expectation is
\begin{equation}\label{eq:leakage}
    \langle \hat\Omega^I_{\kappa}\rangle = \Omega^I_{\kappa} 
    + \sum_{\beta,\kappa'} \Omega^{\beta}_{\kappa'}
    \underbrace{\sum_{\kappa''}
    \bigl(\mathcal{F}^{II}\bigr)^{-1}_{\kappa\kappa''}\,
    \mathcal{F}^{I\beta}_{\kappa''\kappa'}}
    _{\displaystyle \mathcal{L}^{I\beta}_{\kappa\kappa'}}
    \,,
\end{equation}
where $\beta \in \{V,+,-\}$ and $\mathcal{L}^{I\beta}$ is the \textit{leakage matrix}. \black{Analogous leakage matrices arise when the model does not take into account multiple components with different spectral indices \cite{Parida:2015fma, Suresh:2021rsn, DeLillo:2023srz}. For fixed intensity, the leakage scales linearly with polarization fraction, while its magnitude and angular structure depend on the sky morphology and network response.} For persistent sources 
$\Omega^\beta_{\kappa'}$ is fixed, so the bias does not average 
away with longer observation. An exact full block inversion removes this modeled leakage at the cost of larger marginalized intensity uncertainties.

\phantomsection\label{sec:validation}
\textit{Validation.} We simulate a 
persistent, polarized sky observed by the 
LIGO Hanford, LIGO Livingston, and Virgo (HLV) network at expected O5 
sensitivity~\cite{ALIGO} with three years of stationary Gaussian 
noise. The simulation consists of an unpolarized galactic plane 
with amplitude $10^{-6}(f/25\,{\rm Hz})^{2/3}$, a circularly 
polarized cap with polarization degree $p=0.8$ (pure $V$), and a linearly polarized cap 
($p=0.8$, mixed $Q/U$), where $p \equiv \sqrt{Q^2+U^2+V^2}/I$ quantifies the fractional polarization \footnote{See SM Sec.~\ref{sec:map-making} for 
the full parameterization}. We reconstruct the sky in both the SH 
basis (up to maximum multipole $\ell_{\max}=24$) and the pixel basis 
(\textsc{HEALPix} \cite{Gorski:2004by}, $N_{\rm side}=8$, corresponding to $768$ pixels \black{of area $\simeq 54$ deg${}^2$}).

\begin{figure}[htbp]
  \centering
  \includegraphics[width=\columnwidth]{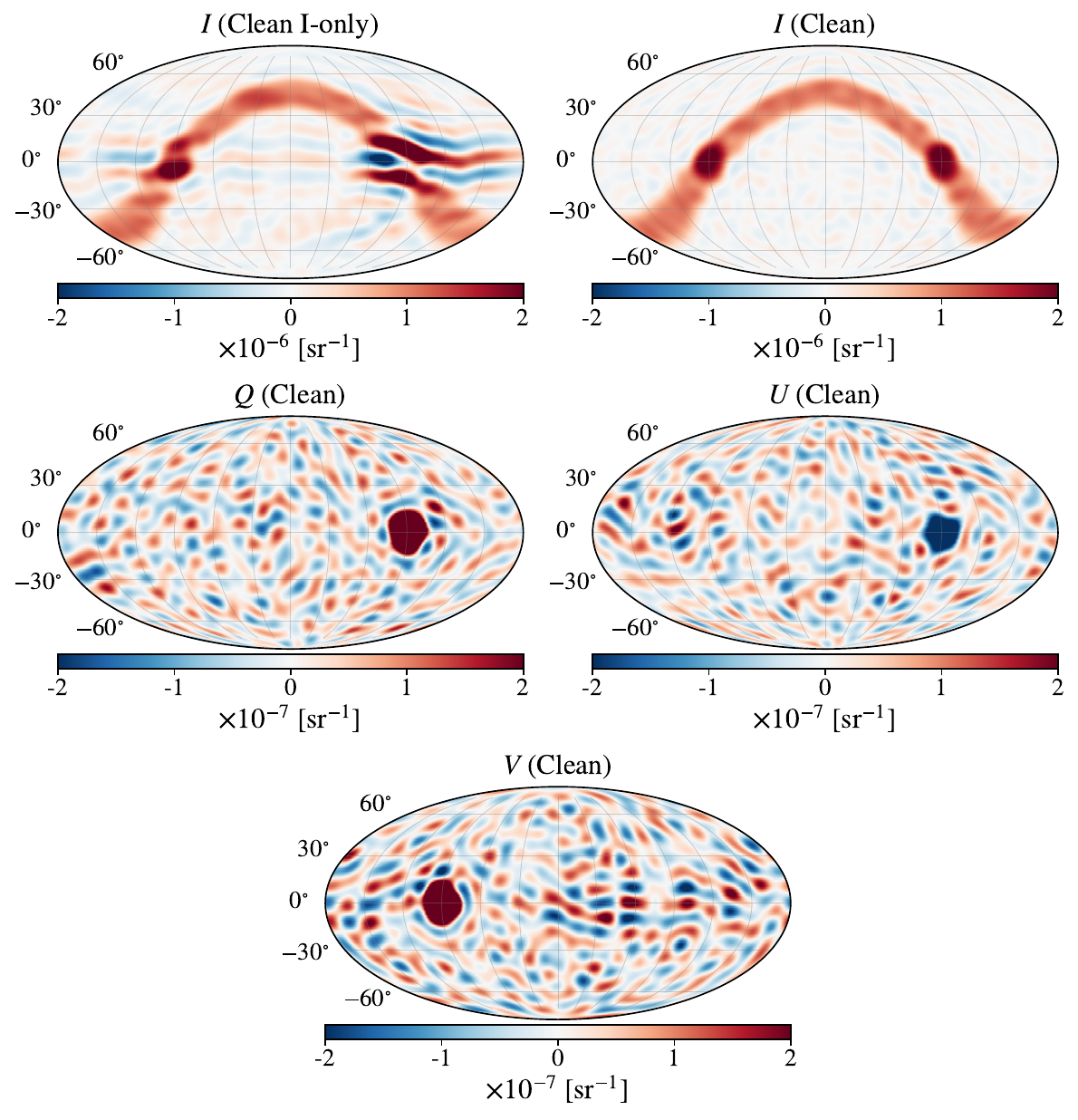}
  \caption{Recovered clean maps in the spherical harmonic basis 
  for the benchmark simulation. Top row compares the $I$ only 
  (left) and full Stokes (right) intensity reconstructions; the 
  $I$ only map shows visible contamination from unmodeled 
  polarization. Bottom row shows the full Stokes $V$, $Q$, and 
  $U$ clean maps, recovering the simulated circularly and linearly 
  polarized caps. Color bars indicate $\Omega_X$ in units of sr$^{-1}$.}
  \label{fig:toy_injection}
\end{figure}

The full Stokes reconstruction recovers the simulated morphology 
in all Stokes sectors in both bases, with per pixel signal to noise ratio
$(\mathrm{SNR})\sim\mathcal{O}(10)$ for the unpolarized plane and 
$\mathrm{SNR}\sim\mathcal{O}(1$\text{--}$10)$ for the polarized caps, as shown in Fig.~\ref{fig:toy_injection}. The normalized residual maps (SM ~\ref{sec:map-making}) are consistent with Gaussian noise, showing no detectable reconstruction bias in the retained modes. Repeating the 
analysis with the $I$-only pipeline, the recovered intensity clean map 
shows visible contamination from the unmodeled polarization
(Fig.~\ref{fig:toy_injection}, top left) and the corresponding normalized residuals exceed $10\sigma$ (SM Fig.~\ref{fig:bias_maps}), directly visualizing the polarization leakage of Eq.~\eqref{eq:leakage}.

\textit{Case study.} In addition to biasing the intensity map, neglecting polarization also leads to the incorrect estimation of the angular power spectrum. Forming $\hat C^{II}_\ell = (2\ell{+}1)^{-1}\sum_m |\hat \Omega^I_{\ell m}|^2$ from the $I$-only clean map and averaging over realizations of the detector noise and the source population gives $\langle \hat C^{II}_\ell \rangle = C^{II}_\ell + \Delta C^{\mathrm{pol}}_\ell + \Delta C^{\mathrm{noise}}_\ell$, where $\Delta C^{\mathrm{noise}}_\ell$ is the detector noise bias~\cite{Kouvatsos:2023bgd} and \black{$\Delta C^{\mathrm{pol}}_\ell$ is the bias coming from the underlying persistent and transient polarized sources (SM Eq. \ref{eq:pol_bias})}. The polarized term is generated by the projection of unmodeled Stokes components through the leakage matrix in Eq.~\eqref{eq:leakage}. For transient CBC foregrounds this contribution has the same observing-time scaling as the ordinary CBC shot noise~\cite{Jenkins:2019uzp,Kouvatsos:2023bgd,Belgacem:2024ohp}, so that it is not parametrically suppressed relative to it.

For compact binaries with isotropically distributed inclinations, the \black{ideal} ratio of the circular-polarization shot-noise power to the intensity shot-noise power is $C^{VV,\mathrm{shot}}_\ell / C^{II,\mathrm{shot}}_\ell = \langle g_V^2 \rangle / \langle g_I^2 \rangle \approx 0.97$, where $g_I = (1+\cos^2\iota)^2/4 + \cos^2\iota$ and $g_V = \cos\iota\,(1+\cos^2\iota)$ \black{describe the inclination dependence of the intensity and of the circular polarization radiated by a binary (SM ~\ref{sec:population})}. Thus polarized shot-noise leakage can be comparable to the conventional CBC intensity shot-noise floor, with the actual observed ratio set by detector geometry and the leakage matrix.

We demonstrate these effects for a third-generation network consisting of one 40 km Cosmic Explorer (CE) at the current LIGO Hanford site~\cite{Reitze:2019iox} and a two-detector Einstein Telescope (ET) in the 2L configuration~\cite{Branchesi:2023mws}, with the two 15 km L-shaped interferometers located at Sos Enattos (Sardinia) and the Meuse-Rhine Euroregion
(Netherlands/Belgium/Germany). \black{The ET detector locations and arm
orientations follow Ref.~\cite{Ebersold:2024hgp} (SM ~\ref{sec:population})}. We simulate three
years of observation and model the astrophysical foreground using a
CBC population\textcolor{black}{~\cite{Regimbau:2008nj,Zhu:2011bd,Wu:2011ac,Zhu:2012xw}} calibrated to \black{GWTC-5}
rates~\cite{LIGOScientific:2026ctl} (SM ~\ref{sec:population}),
with events above a network $\mathrm{SNR} > 20$ individually
resolved and assumed to be known. \black{The residual population forms an
anisotropic polarized shot-noise foreground (SM
Fig.\ref{fig:clbands_pop}). On top of this foreground we simulate a
circularly polarized cosmic monopole with $\alpha=0$ with
$\Omega_I^{\rm cos}=10^{-11}$ and $\Pi_V \equiv
\Omega_V^{\mathrm{cos}}/\Omega_I^{\mathrm{cos}} = 0.1$,
representative of a primordial chiral signal. Such a component
could arise from axion-gauge inflation, whose chiral spectrum can
rise into the ground-based
band~\cite{Cook:2011hg,Crowder:2012ik,Domcke:2019zls}, or from the
chirality of scalar-induced GWs~\cite{Caporali:2026qhc}. We then
compare the current searches that target either the intensity
anisotropies $\hat{C}^{II}_{\ell}$
\cite{LIGOScientific:2025kbu,LIGOScientific:2019gaw,LIGOScientific:2016nwa}
or the parity violating monopole $\hat{\Omega}^V$
\cite{LIGOScientific:2025kry} against the full-Stokes search which fits all Stokes multipoles jointly. }

\begin{figure}[htbp]
  \centering
  \includegraphics[width=\columnwidth]{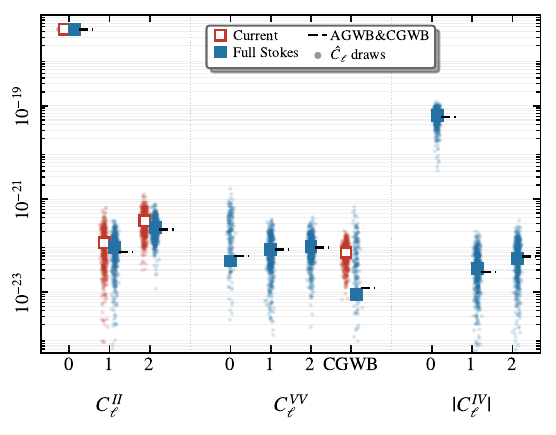}
  \caption{Angular power spectra for a CE+ET(2L) network with a simulated CBC foreground and a circularly polarized cosmic monopole ($\Pi_V = 0.1$, $\alpha = 0$). Open (filled) markers show the $I$-only (full-Stokes) analysis. Dashes mark \black{the total astrophysical (AGWB) and cosmological (CGWB) background and the dots show individual Monte Carlo realizations.} }
  \label{fig:3g_forecast}
\end{figure}

\black{In Fig.~\ref{fig:3g_forecast}, we show the recovery of the total background over 100 realizations for the current searches and the full-Stokes case.  Both analyses recover the monopole
$\hat\Omega^I_{00}$ with a small fractional bias, because the
leaked polarization is a small addition to the large total
intensity. For the $\ell>0$ multipoles the polarized shot noise is
as large as the intensity shot noise itself, and the $I$-only
spectrum sits above the full-Stokes one by the predicted $\Delta
C^{\mathrm{pol}}_\ell$. Since $\mathcal{F}^{I\beta}_{\kappa\kappa'}
= \mathcal{F}^{\beta I\,*}_{\kappa\kappa'} \neq 0$, the intensity
shot noise in turn biases the recovered circular polarization, so a parity violating signal cannot be claimed with
the current searches in the presence of this foreground.}

The full-Stokes fit instead leaves the $\ell>0$ residuals at the
expected shot-noise level, recovers the cosmic $V$ amplitude, and
with it the polarization fraction $\Pi_V$, and gives direct access
to the parity-odd cross spectra $C^{IV}_\ell$. The cost of the
joint fit is a factor of about $1.4$ in the detector-noise
contribution to the monopole variance (SM Sec.~\ref{sec:cov}),
while the $I$-only estimator carries additional
realization-dependent variance from the leakage across different modes, so the total scatters of the two
analyses remain comparable. \black{We also show that these
conclusions are robust against variations of the detection
threshold, the population assumptions, the analysis $\ell_{\max}$,
the regularization cutoff and the observation time, and longer
observation only raises the significance of the bias (SM
Sec.~\ref{sec:population}). We note that these effects are not specific to the case
study. The isotropic intensity generally dominates, while anisotropies and polarizations of transient or persistent sources can be comparable and couple through the detector response, so a joint analysis is needed when the leakage term in Eq.\eqref{eq:leakage} becomes comparable to the statistical uncertainty.}

\begin{figure*}[htbp]
  \centering
  \includegraphics[width=0.24\textwidth]{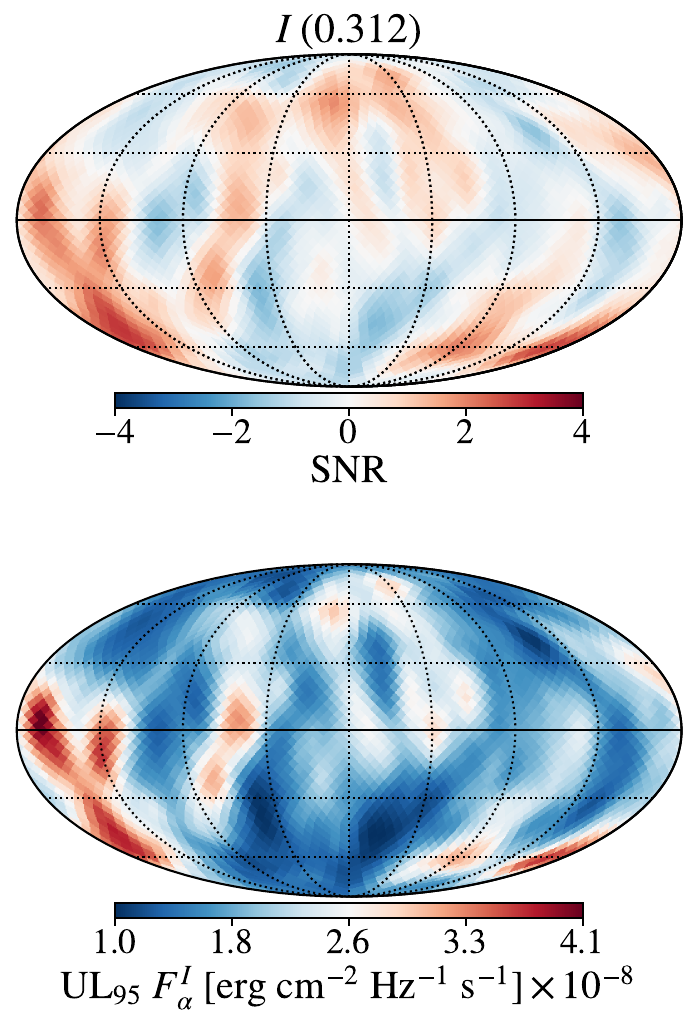}
  \hfill
  \includegraphics[width=0.24\textwidth]{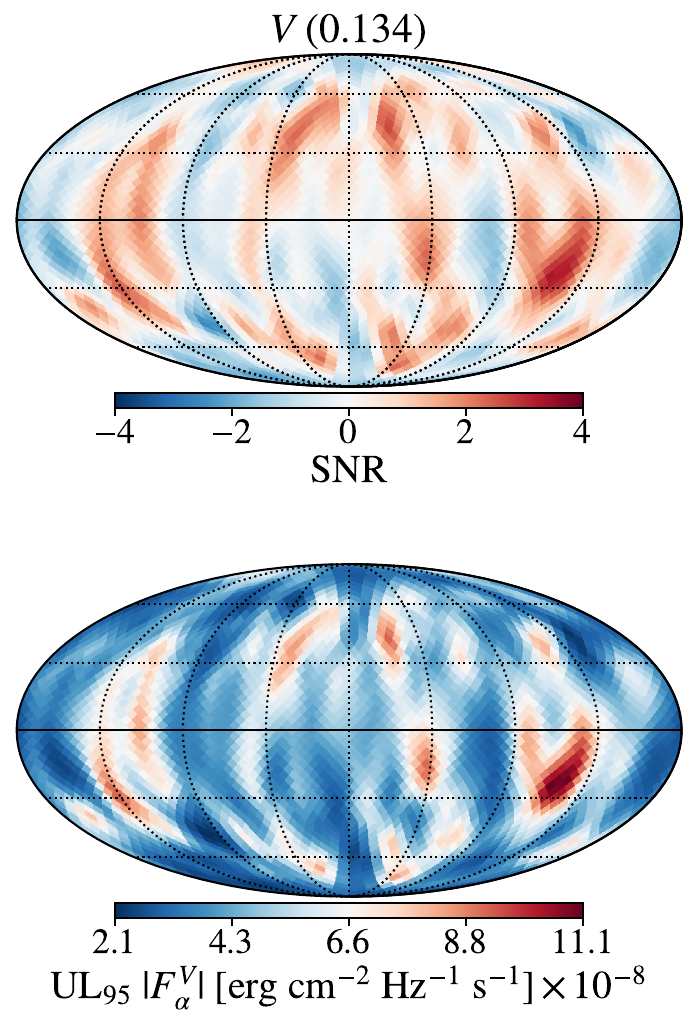}
  \hfill
  \includegraphics[width=0.24\textwidth]{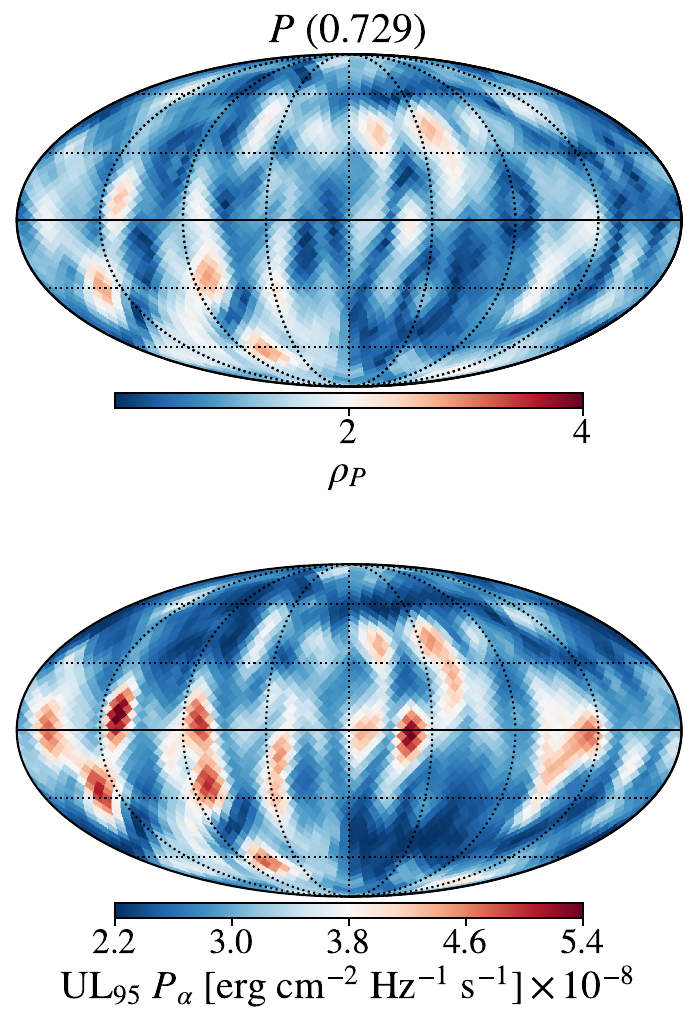}
  \hfill
  \includegraphics[width=0.24\textwidth]{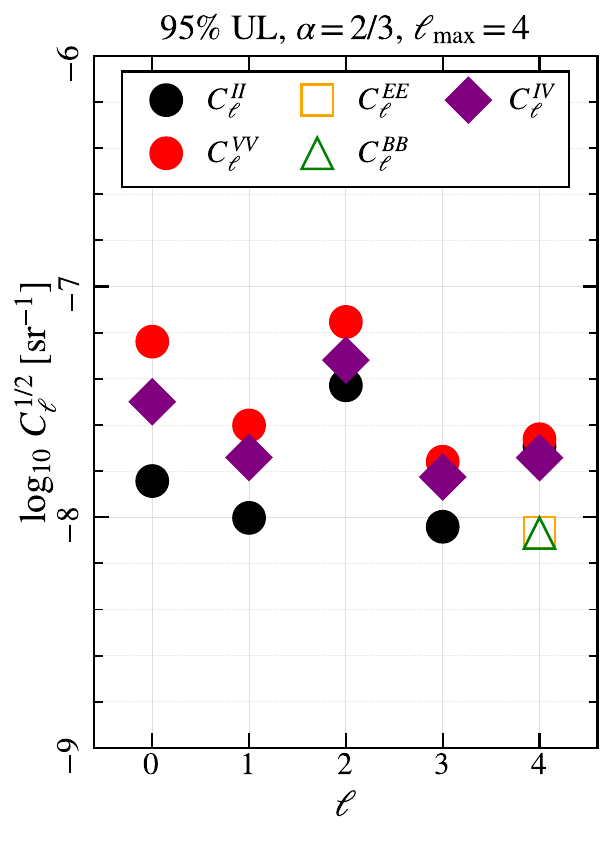}
  \caption{Results from the full-Stokes analysis of the LVK O3 data at $\alpha=2/3$. The first three columns show the SNR (top) and the 95\% upper limit (bottom) maps for Intensity ($I$), Circular Polarization ($V$), and total Linear Polarization amplitude ($P \equiv \sqrt{Q^2+U^2}$), respectively. Numbers in parentheses in the map titles give the corresponding global
 $p$-values (SM Sec. \ref{sec:o3_extra}). The fourth panel displays the corresponding 95\% bounds on the angular spectra. For the parity-odd cross spectrum, we plot the 95\% bound $|C^{IV}_\ell|_{95}^{1/2}$. Since $E$ and $B$ begin at $\ell=4$, the O3 linear-polarization power constraints correspond only to $\ell=4$.}
  \label{fig:o3_wide_results}
\end{figure*}

\textit{O3 results.} We apply the full Stokes formalism to the publicly available folded cross-correlation data from the first three observing runs of Advanced LIGO and Advanced Virgo~\cite{KAGRA:2021mth,KAGRA:2021kbb}, the most recent dataset for which the required baseline cross-spectra are publicly released~\cite{LVK:O3FoldedData,LIGOScientific:2025kbu}. Pixel basis upper limit maps are reported as GW energy flux density $F_\alpha$ (SM Eq.~\eqref{eq:flux_def}); angular power spectra are reported as $\Omega$. We combine the HL baseline from O1, O2, and O3, together with the HV and LV baselines from O3. In the pixel basis ($N_{\mathrm{side}} = 16$, corresponding to $3072$ pixels \black{of area $\simeq 13$ deg${}^2$)} we produce sky maps for spectral indices $\alpha \in \{0, 2/3, 3\}$; in the spherical harmonic basis we analyze up to $\ell_{\mathrm{max}} = 4$ at $\alpha = 2/3$, following the pixel and spherical-harmonic map-making approaches used in analyses~\cite{Ain:2018zvo,Suresh:2020khz,KAGRA:2021mth}. Since $E$ and $B$ begin at $\ell=4$, this O3 analysis constrains only the $\ell=4$ linear-polarization power. From the clean maps we estimate the angular 
power spectra~\cite{Thrane_2009}
\begin{equation}\label{eq:cl_est}
    \hat C^{XY}_\ell = \frac{1}{2\ell+1}\sum_m 
    \mathrm{Re}[\hat\Omega^X_{\ell m}\,
    \hat\Omega^{Y*}_{\ell m}] 
    - N^{XY}_\ell\,,
\end{equation}
where $N^{XY}_\ell = (2\ell+1)^{-1}\sum_m 
\mathrm{Re}[(\mathcal{F}^{-1})^{XY}_{\ell m, \ell m}]$ is the corresponding noise bias.

The results at $\alpha = 2/3$ are summarized in Fig.~\ref{fig:o3_wide_results} (see SM Fig.~\ref{fig:o3_extra} for $\alpha=0$ and $\alpha=3$). 
We find no significant anisotropy in any Stokes parameter. Each map is characterized by the per-pixel radiometer SNR map, $\hat X_{\hat n}/\sqrt{\mathcal{F}_{\hat n\hat n}}$~\cite{Thrane_2009,KAGRA:2021mth}; for linear polarization we use the statistic
\(\rho_P^2(p)=(\hat\Omega_p^Q,\hat\Omega_p^U)
\left([(\mathcal F^{-1})_{pp}^{XY}]_{X,Y=Q,U}\right)^{-1}
(\hat\Omega_p^Q,\hat\Omega_p^U)^T\) that is invariant under polarization basis rotation \black{\cite{QUIET:2014szn}}. The global significance is assessed from a noise-only Monte Carlo that preserves the sky covariance, detailed in SM Sec.~\ref{sec:o3_methods}. At $\alpha=2/3$ the maximum per-pixel values are $\mathrm{SNR}\simeq2.9$ for $I$, $3.2$ for $V$ and $\rho_P\simeq2.8$, with global $p$-values $0.31$, $0.13$ and $0.73$; the corresponding null distributions are shown in SM Fig.~\ref{fig:o3_pix_pval}. The first three columns of Fig.~\ref{fig:o3_wide_results} show the SNR and the $95\%$ upper-limit maps for $I$, $V$ and the total linear-polarization amplitude $P\equiv\sqrt{Q^2+U^2}$. The fourth panel presents
the $95\%$ bounds on the angular spectra, with the calibration-marginalized Bayesian posterior of Ref.~\cite{Whelan:2012zw}. \black{Marginalizing over the polarized Stokes parameters broadens the
intensity bound, with the joint median $C^{II}_\ell{}^{1/2}$ upper limits
a factor of $2$--$3$ above an $I$-only analysis and the
pixel-basis limits within $\sim10\%$ of the published
intensity-only maps~\cite{KAGRA:2021mth} (SM
Fig.~\ref{fig:lvk_vs_full}). The $C^{VV}_0$ bound corresponds to $\Omega_V \simeq 1.6\times10^{-8}$ at $\alpha=2/3$, within a factor of a few of the dedicated isotropic parity searches~\cite{Martinovic:2021hzy} (SM Sec.~\ref{sec:o3_extra}). Among the mechanisms of the case study, maximally chiral axion-gauge inflation can rise to $\Omega_V\sim10^{-9}$ within its allowed parameter space~\cite{Barnaby:2011qe,Crowder:2012ik}, while the scalar-induced and astrophysical chiralities lie within reach of the next-generation networks~\cite{Caporali:2026qhc,ValbusaDallArmi:2023ydl}.}

The parity-odd cross-spectrum $C^{IV}_\ell$ is of particular
interest as a probe of gravitational parity violation. For an
isotropically oriented CBC population, the inclination-averaged
cross-response vanishes ($\langle g_I\,g_V\rangle = 0$; SM
Sec.~\ref{sec:population}), so the ensemble-mean CBC shot-noise bias
in $C^{IV}_\ell$ vanishes. Finite realizations still contribute
variance, but not a positive semidefinite systematic bias analogous
to $C^{VV}_\ell$.

\black{Our results could prove useful in several directions. The joint fit
turns the $\Omega_V$ bound into a reliable estimate of gravitational
parity violation, from mechanisms such as Chern-Simons
gravity~\cite{Alexander:2009tp}, chiral production in the early
Universe~\cite{Seto:2008sr,Callister:2017ocg} or birefringent
propagation~\cite{Martinovic:2021hzy,Callister:2023tws}, since
intensity foreground content leaks into the $V$ channel unless
jointly modeled (SM Sec.~\ref{sec:population}). For a
parity-symmetric background the cross spectra $C^{IV}_\ell$ and
$C^{EB}_\ell$ vanish, while for isotropically oriented CBC populations the ensemble-mean linear polarization vanishes, providing useful null tests. The $V$ channel also
characterizes the CBC shot noise, whose circular-polarization power
is comparable to the intensity shot noise (SM
Sec.~\ref{sec:population}). An unmodeled polarized persistent
component biases the intensity-only estimator, as the benchmark
simulation demonstrates (Fig.~\ref{fig:toy_injection}), and is
corrected by the joint treatment (SM
Sec.~\ref{sec:bias_derivation}). The polarized maps could also
extend cross-correlations with other cosmological tracers beyond the
intensity channel~\cite{Cusin:2017fwz,Jenkins:2018nty}.}

\textit{Conclusion.} We have implemented a full Stokes SGWB search framework that jointly reconstructs $I,V,Q,U$ from existing ground-based cross-correlation data. Applied to LVK O3 data, we find no significant polarization anisotropy and place bounds on $C^{VV}_\ell$, $C^{EE}_\ell$, $C^{BB}_\ell$, and $|C^{IV}_\ell|$. Simulations for future detector networks demonstrate that neglecting polarization can bias intensity-only anisotropy estimates through polarized sky components and CBC shot-noise leakage. The full Stokes framework separates the Stokes sectors and provides access to polarized anisotropies invisible to conventional intensity-only searches.

\textit{Acknowledgments:} \blue{The data products of this work are publicly available on Zenodo~\cite{Boybeyi:2026zenodo}.}  We are grateful to Deepali Agarwal and the LVK Stochastic Group for valuable comments and feedback on the draft. TB and VM were in part supported by the NSF grant PHY-2409173. 
This research has made use of data or software obtained from the Gravitational Wave Open Science Center (gwosc.org), a service of the LIGO Scientific Collaboration, the Virgo Collaboration, and KAGRA. This material is based upon work supported by LIGO Laboratory which is a major facility fully funded by the National Science Foundation, as well as the Science and Technology Facilities Council (STFC) of the United Kingdom, the Max-Planck-Society (MPS), and the State of Niedersachsen/Germany for support of the construction of Advanced LIGO and construction and operation of the GEO600 detector. Additional support for Advanced LIGO was provided by the Australian Research Council. Virgo is funded, through the European Gravitational Observatory (EGO), by the French Centre National de Recherche Scientifique (CNRS), the Italian Istituto Nazionale di Fisica Nucleare (INFN) and the Dutch Nikhef, with contributions by institutions from Belgium, Germany, Greece, Hungary, Ireland, Japan, Monaco, Poland, Portugal, Spain. KAGRA is supported by Ministry of Education, Culture, Sports, Science and Technology (MEXT), Japan Society for the Promotion of Science (JSPS) in Japan; National Research Foundation (NRF) and Ministry of Science and ICT (MSIT) in Korea; Academia Sinica (AS) and National Science and Technology Council (NSTC) in Taiwan.

\bibliography{apssamp}
\clearpage
\onecolumngrid

\setcounter{equation}{0}   
\setcounter{figure}{0}    
\setcounter{table}{0}     
\setcounter{section}{0} 
\setcounter{secnumdepth}{2}
\renewcommand{\theequation}{S\arabic{equation}}  
\renewcommand{\thefigure}{S\arabic{figure}}     
\renewcommand{\thetable}{S\arabic{table}}      
\renewcommand{\thesection}{S-\Roman{section}}    
\begin{center}
\textbf{\large Supplemental Material (SM)}
\end{center}

\section{General Definitions} \label{sec:conventions}

Our orthonormal triad, with $\nhat$ depicting the \black{sky} direction of an incoming gravitational wave, is defined as
\begin{align}
\nhat(\theta,\phi)
&=(\sin\theta\cos\phi,\ \sin\theta\sin\phi,\ \cos\theta),\\
\phat(\theta,\phi)
&=(\cos\theta\cos\phi,\ \cos\theta\sin\phi,\ -\sin\theta),\\
\qhat(\theta,\phi)
&=(-\sin\phi,\ \cos\phi,\ 0),
\end{align}
where $\theta,\phi$ are the usual spherical coordinates. We adopt the following decomposition of a generic 2-index tensor:
\begin{align}
e^{(+2)}_{ij}(\nhat)
&=\frac{1}{\sqrt2}\,(\phat_i+i\qhat_i)(\phat_j+i\qhat_j),&
e^{(-2)}_{ij}(\nhat)
&=\frac{1}{\sqrt2}\,(\phat_i-i\qhat_i)(\phat_j-i\qhat_j),\\
e^{(+1)}_{ij}(\nhat)
&=-\frac{1}{\sqrt2}\Big[(\phat_i+i\qhat_i)\nhat_j+\nhat_i(\phat_j+i\qhat_j)\Big],&
e^{(-1)}_{ij}(\nhat)
&=-\frac{1}{\sqrt2}\Big[(\phat_i-i\qhat_i)\nhat_j+\nhat_i(\phat_j-i\qhat_j)\Big],\\
e^{(+0)}_{ij}(\nhat)
&=\phat_i\phat_j+\qhat_i\qhat_j,&
e^{(-0)}_{ij}(\nhat)
&=\sqrt{2}\,\nhat_i\nhat_j.
\end{align}
such that $\tlh_{ij}(f,\nhat)=\sum_{s\in\{\pm 2,\pm 1, \pm 0\}} e^{(s)}_{ij}(\nhat)\,\tlh^{(s)}(f,\nhat)$.

Rotation of the triad with the form:
\begin{align}
\phat(\psi) &= \cos\psi\,\phat + \sin\psi\,\qhat,\\
\qhat(\psi) &= -\sin\psi\,\phat + \cos\psi\,\qhat
\end{align}
gives
\begin{align}
e^{(s)}_{ij}(\nhat,\psi)=e^{-is\psi}\,e^{(s)}_{ij}(\nhat),
\qquad s\in\{\pm 2,\pm 1, \pm 0\},
\end{align}
hence justifies the labeling as spin/helicity-weight.

We express the detector output in the frequency domain for a time segment labeled by $t$ as a sum of the GW signal and noise contributions:
\begin{align}
\tilde{s}_{\mathcal I}(f;t)=\tilde{h}_{\mathcal I}(f;t)+\tilde{n}_{\mathcal I}(f;t).
\end{align}
For a plane wave,
\begin{align}
h_{ij}(t,\vec x)=\int_{-\infty}^{\infty} df \int d\Omega\; \tilde{h}_{ij}(f,\nhat)\,e^{i2\pi f\,(t+\nhat\cdot \vec x/c)}.
\end{align}
In the long-wavelength approximation for a ground-based interferometer (limiting to tensor polarizations),
\begin{align}
\tilde{h}_{\mathcal I}(f;t)
&=\sum_{A=+,\times}\int_{S^{2}} d\Omega\;F^{A}_{\mathcal I}(\nhat,t)\,\tilde{h}_{A}(f,\nhat)\,
e^{i\frac{2\pi f}{c}\,\nhat\cdot \vec x_{\mathcal I}(t)},
\end{align}
where $F^{A}_{\mathcal I}(\nhat;t)$ are the antenna pattern functions for polarization $A \in \{+,\times\}$ and
\begin{align}
\tilde{h}_{ij}(f,\nhat)=\sum_{A=+,\times} e^{A}_{ij}(\nhat)\,\tilde{h}_{A}(f,\nhat).
\end{align}
In this approximation the $f$-dependence of the full response comes only from the phase factor $e^{i(2\pi f/c)\nhat\cdot \vec x_{\mathcal I}}$ (i.e.\ $F^A_{\mathcal I}$ is independent of $f$).

For the stochastic background polarization correlations we adopt the Stokes-parameter matrix in $+/\times$ basis \cite{Romano:2016dpx}. Note that the main text Eqs.~\eqref{eq:GW_strain_correlations_helicity}--\eqref{eq:H_helicity_matrix} use the helicity ($\pm2$) basis, while here we work in the linear ($+/\times$) basis; the two are related by a unitary rotation, and the Stokes parameters $\{I,Q,U,V\}$ are identical in both representations:
\begin{align}
\Big\langle \tilde{h}_{A}(f,\nhat)\,\tilde{h}^{*}_{A'}(f',\nhat')\Big\rangle
=\frac{1}{2}\,\delta(f-f')\,\delta^{(2)}(\nhat-\nhat')\,S^{AA'}_{h}(f,\nhat),
\end{align}
with
\begin{align}
S^{AA'}_{h}(f,\nhat)
=\frac{1}{2}
\begin{pmatrix}
I(f,\nhat)+Q(f,\nhat) & U(f,\nhat)-iV(f,\nhat)\\
U(f,\nhat)+iV(f,\nhat) & I(f,\nhat)-Q(f,\nhat)
\end{pmatrix}_{AA'}.
\end{align}

Using \black{the detector response and the Stokes-parameter matrix above}, and approximating the finite-time delta as $\delta(0)\rightarrow \tau$ for a segment of duration $\tau$, the signal cross-correlation at equal frequency is
\begin{align}
\Big\langle \tilde{h}_{\mathcal I}(f;t)\,\tilde{h}^{*}_{\mathcal J}(f;t)\Big\rangle
&=\frac{\tau}{4}\int d\Omega\;
e^{i\frac{2\pi f}{c}\,\nhat\cdot \Delta \vec x_{\mathcal I\mathcal J}(t)}
\Big[
\big(F^+_{\mathcal I}F^+_{\mathcal J}+F^\times_{\mathcal I}F^\times_{\mathcal J}\big) I
+\big(F^+_{\mathcal I}F^+_{\mathcal J}-F^\times_{\mathcal I}F^\times_{\mathcal J}\big) Q \nonumber\\
&\hspace{4.2cm}
+\big(F^+_{\mathcal I}F^\times_{\mathcal J}+F^\times_{\mathcal I}F^+_{\mathcal J}\big) U
-i\big(F^+_{\mathcal I}F^\times_{\mathcal J}-F^\times_{\mathcal I}F^+_{\mathcal J}\big) V
\Big],
\end{align}
where all Stokes parameters and antenna patterns are evaluated at $(f,\nhat)$ and
\begin{align}
\Delta \vec x_{\mathcal I\mathcal J}(t)\equiv \vec x_{\mathcal I}(t)-\vec x_{\mathcal J}(t).
\end{align}

It is convenient to define direction-dependent overlap reduction functions (evaluated at a reference time $t$ over a short segment):
\begin{align}
\Gamma^{I}_{\mathcal I\mathcal J}(f,\nhat;t)
&\equiv \frac{1}{2}\Big(F^+_{\mathcal I}(\nhat,t)\,F^+_{\mathcal J}(\nhat,t)+F^\times_{\mathcal I}(\nhat,t)\,F^\times_{\mathcal J}(\nhat,t)\Big)
e^{i\frac{2\pi f}{c}\,\nhat\cdot \Delta \vec x_{\mathcal I\mathcal J}(t)},\\
\Gamma^{Q}_{\mathcal I\mathcal J}(f,\nhat;t)
&\equiv \frac{1}{2}\Big(F^+_{\mathcal I}(\nhat,t)\,F^+_{\mathcal J}(\nhat,t)-F^\times_{\mathcal I}(\nhat,t)\,F^\times_{\mathcal J}(\nhat,t)\Big)
e^{i\frac{2\pi f}{c}\,\nhat\cdot \Delta \vec x_{\mathcal I\mathcal J}(t)},\\
\Gamma^{U}_{\mathcal I\mathcal J}(f,\nhat;t)
&\equiv \frac{1}{2}\Big(F^+_{\mathcal I}(\nhat,t)\,F^\times_{\mathcal J}(\nhat,t)+F^\times_{\mathcal I}(\nhat,t)\,F^+_{\mathcal J}(\nhat,t)\Big)
e^{i\frac{2\pi f}{c}\,\nhat\cdot \Delta \vec x_{\mathcal I\mathcal J}(t)},\\
\Gamma^{V}_{\mathcal I\mathcal J}(f,\nhat;t)
&\equiv -\frac{i}{2}\Big(F^+_{\mathcal I}(\nhat,t)\,F^\times_{\mathcal J}(\nhat,t)-F^\times_{\mathcal I}(\nhat,t)\,F^+_{\mathcal J}(\nhat,t)\Big)
e^{i\frac{2\pi f}{c}\,\nhat\cdot \Delta \vec x_{\mathcal I\mathcal J}(t)}.
\end{align}
Define also the spin-$\pm 4$ combinations
\begin{align}
P^{\pm}(f,\nhat)\equiv Q(f,\nhat)\pm iU(f,\nhat),
\qquad
\Gamma^{\pm}_{\mathcal I\mathcal J}(f,\nhat;t)\equiv \frac{1}{2}\Big(\Gamma^{Q}_{\mathcal I\mathcal J}(f,\nhat;t)\mp i\,\Gamma^{U}_{\mathcal I\mathcal J}(f,\nhat;t)\Big),
\end{align}
so that $\Gamma^{Q}Q+\Gamma^{U}U=\Gamma^{+}P^{+}+\Gamma^{-}P^{-}$.

Spin-weighted spherical harmonics are defined by
\begin{align}
{}_{s}Y_{\ell m}(\theta,\phi)
= (-1)^s \sqrt{\frac{2\ell+1}{4\pi}}\; D^{\,\ell*}_{m,\,-s}(\phi,\theta,0),
\qquad \ell \ge |s|,\ \ |m|\le \ell,
\end{align}
with Wigner $D$-matrices
\begin{align}
D^{\ell}_{m m'}(\alpha,\beta,\gamma)
= e^{-i m \alpha}\, d^{\ell}_{m m'}(\beta)\, e^{-i m' \gamma},
\end{align}
\black{where $d^{\ell}_{m m'}(\beta)$'s are defined in \cite{Sakurai_Napolitano_2020}} and normalization
\begin{align}
\int_{S^{2}} d\Omega\ {}_{s}Y_{\ell m}(\nhat)\,{}_{s}Y^{*}_{\ell'm'}(\nhat)\,
= \delta_{\ell \ell'}\,\delta_{mm'}.
\end{align}

We expand each Stokes field in spin-weighted spherical harmonics, writing $\Omega_X(\nhat) = \sum_{\ell m} a^X_{\ell m}\,{}_{s_X}\!Y_{\ell m}(\nhat)$ with $s_X = 0$ for $X \in \{I,V\}$ and $s_X = \pm4$ for $X = P^\pm$ (so that $a^X_{\ell m} \equiv \Omega^X_{\ell m}$ in the notation of Eq.~\eqref{eq:stokes_harm}).

The gradient ($E$) and curl ($B$) components of the linear polarization are
\begin{align}
    a^E_{\ell m} &= -\tfrac{1}{2}(a^+_{\ell m} + a^-_{\ell m}), &
    a^B_{\ell m} &= \tfrac{i}{2}(a^+_{\ell m} - a^-_{\ell m}), \label{eq:EB_def}
\end{align}
with inverse
\begin{align}
    a^+_{\ell m} &= -a^E_{\ell m} - i\,a^B_{\ell m}, &
    a^-_{\ell m} &= -a^E_{\ell m} + i\,a^B_{\ell m}. \label{eq:EB_inv}
\end{align}
Under parity $\nhat \to -\nhat$, $E$ is even and $B$ is odd. The angular power spectra are
\begin{equation}\label{eq:cl_def_sm}
    C^{XY}_\ell = \frac{1}{2\ell+1}\sum_{m=-\ell}^{\ell} \mathrm{Re}\!\left[a^X_{\ell m}\,a^{Y*}_{\ell m}\right], \qquad X,Y \in \{I,V,E,B\}.
\end{equation}
For a parity-symmetric background $C^{IV}_\ell = C^{IB}_\ell = C^{VE}_\ell = C^{EB}_\ell = 0$.

The corresponding ORFs are
\begin{align}
\Gamma^{I}_{\mathcal I\mathcal J,\ell m}(f,t)
&\equiv \int_{S^{2}} d\Omega\; Y_{\ell m}(\nhat)\,\Gamma^{I}_{\mathcal I\mathcal J}(f,\nhat;t),\\
\Gamma^{V}_{\mathcal I\mathcal J,\ell m}(f,t)
&\equiv \int_{S^{2}} d\Omega\; Y_{\ell m}(\nhat)\,\Gamma^{V}_{\mathcal I\mathcal J}(f,\nhat;t),\\
\Gamma^{+}_{\mathcal I\mathcal J,\ell m}(f,t)
&\equiv \int_{S^{2}} d\Omega\; {}_{+4}Y_{\ell m}(\nhat)\,\Gamma^{+}_{\mathcal I\mathcal J}(f,\nhat;t),\\
\Gamma^{-}_{\mathcal I\mathcal J,\ell m}(f,t)
&\equiv \int_{S^{2}} d\Omega\; {}_{-4}Y_{\ell m}(\nhat)\,\Gamma^{-}_{\mathcal I\mathcal J}(f,\nhat;t).
\end{align}

Combining the above and using orthonormality gives
\begin{align}
&\Big\langle \tilde{h}_{\mathcal I}(f;t)\,\tilde{h}^{*}_{\mathcal J}(f;t)\Big\rangle
=
\tau\,\frac{3H_0^2}{4\pi^2 f^3}\,H(f)
\sum_{\ell m}\Big[
\Gamma^{I}_{\mathcal I\mathcal J,\ell m}(f,t)\,a^{I}_{\ell m}
+\Gamma^{V}_{\mathcal I\mathcal J,\ell m}(f,t)\,a^{V}_{\ell m}
\nonumber\\
&\hspace{4.5cm}
+\Gamma^{+}_{\mathcal I\mathcal J,\ell m}(f,t)\,a^{+}_{\ell m}
+\Gamma^{-}_{\mathcal I\mathcal J,\ell m}(f,t)\,a^{-}_{\ell m}
\Big].
\end{align}
Equivalently, for $\hat C_{\mathcal I\mathcal J}(f;t)=\tau^{-1}\tilde s_{\mathcal I}(f;t)\tilde s^*_{\mathcal J}(f;t)$,
\begin{equation}
\Big\langle \hat C_{\mathcal I\mathcal J}(f;t)\Big\rangle
=
\sum_{\ell m,X}
\tilde\Gamma^{X}_{\mathcal I\mathcal J,\ell m}(f,t)\,
a^X_{\ell m},
\end{equation}
with $\tilde\Gamma^{X}_{\mathcal I\mathcal J,\ell m}$ defined as the $\Omega$-normalized response in the main text.

\begin{figure*}[!htbp]
  \centering
  \begin{minipage}[b]{0.48\textwidth}\centering
    \includegraphics[width=\linewidth]{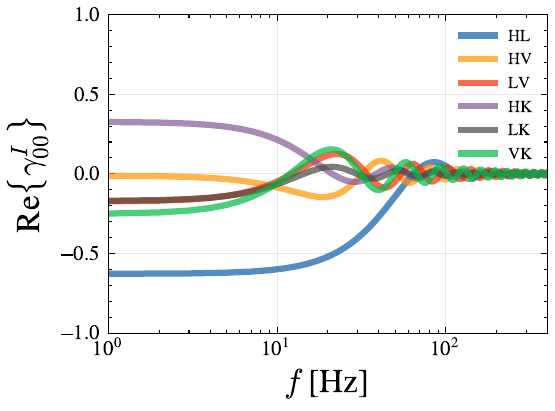}\\[-2pt]
    \scriptsize $I$ overlap--reduction functions for $\ell =0,\, m=0$.
  \end{minipage}\hfill
  \begin{minipage}[b]{0.48\textwidth}\centering
    \includegraphics[width=\linewidth]{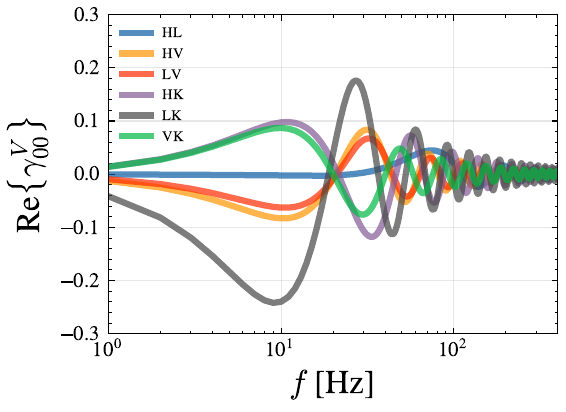}\\[-2pt]
    \scriptsize $V$ overlap--reduction functions for $\ell =0,\, m=0$.
  \end{minipage}

  \vspace{2mm}
  \begin{minipage}[b]{0.48\textwidth}\centering
    \includegraphics[width=\linewidth]{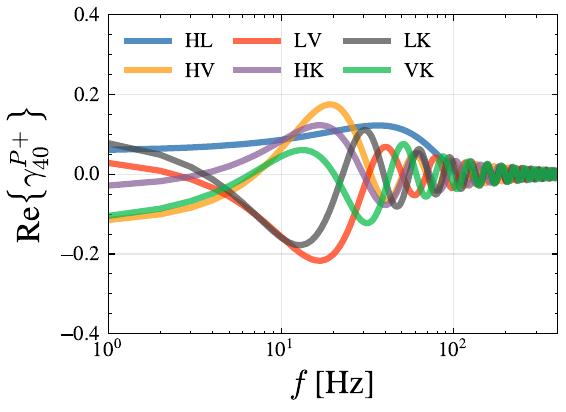}\\[-2pt]
    \scriptsize $P_{+}$ overlap–reduction functions for $\ell =4$.
  \end{minipage}\hfill
  \begin{minipage}[b]{0.48\textwidth}\centering
    \includegraphics[width=\linewidth]{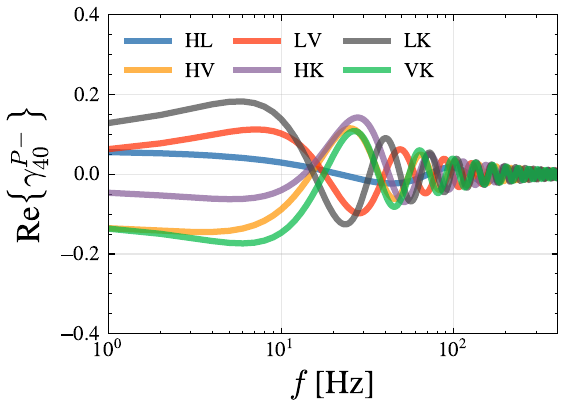}\\[-2pt]
    \scriptsize $P_{-}$ overlap–reduction functions for $\ell =4$.
  \end{minipage}

  \caption{Real parts of the ORFs $\gamma^{X}_{\ell m}(f) = 1/\sqrt{4\pi } \ \Gamma^{X}_{\ell m}(f) $ normalized with the spherical harmonics conventions in \cite{Romano:2016dpx} (Fig. 45) for selected modes and polarizations. Imaginary parts of ORF's vanish for $m=0$.}
  \label{fig:orf_grid}
\end{figure*}

Exemplary ORFs are given in Fig. \ref{fig:orf_grid}. The sidereal phase factor $e^{im\Omega_\oplus(t-t_0)}$ used for data folding (Sec.~\ref{sec:map-making}) is conventional and we adopt the \cite{Romano:2016dpx} conventions. Also, we do not attempt to provide analytical expressions for these functions; for a derivation see \cite{Chu:2020qiw,Liu:2022umx}.

\begin{figure*}[htbp]
  \centering
  \includegraphics[width=\textwidth]{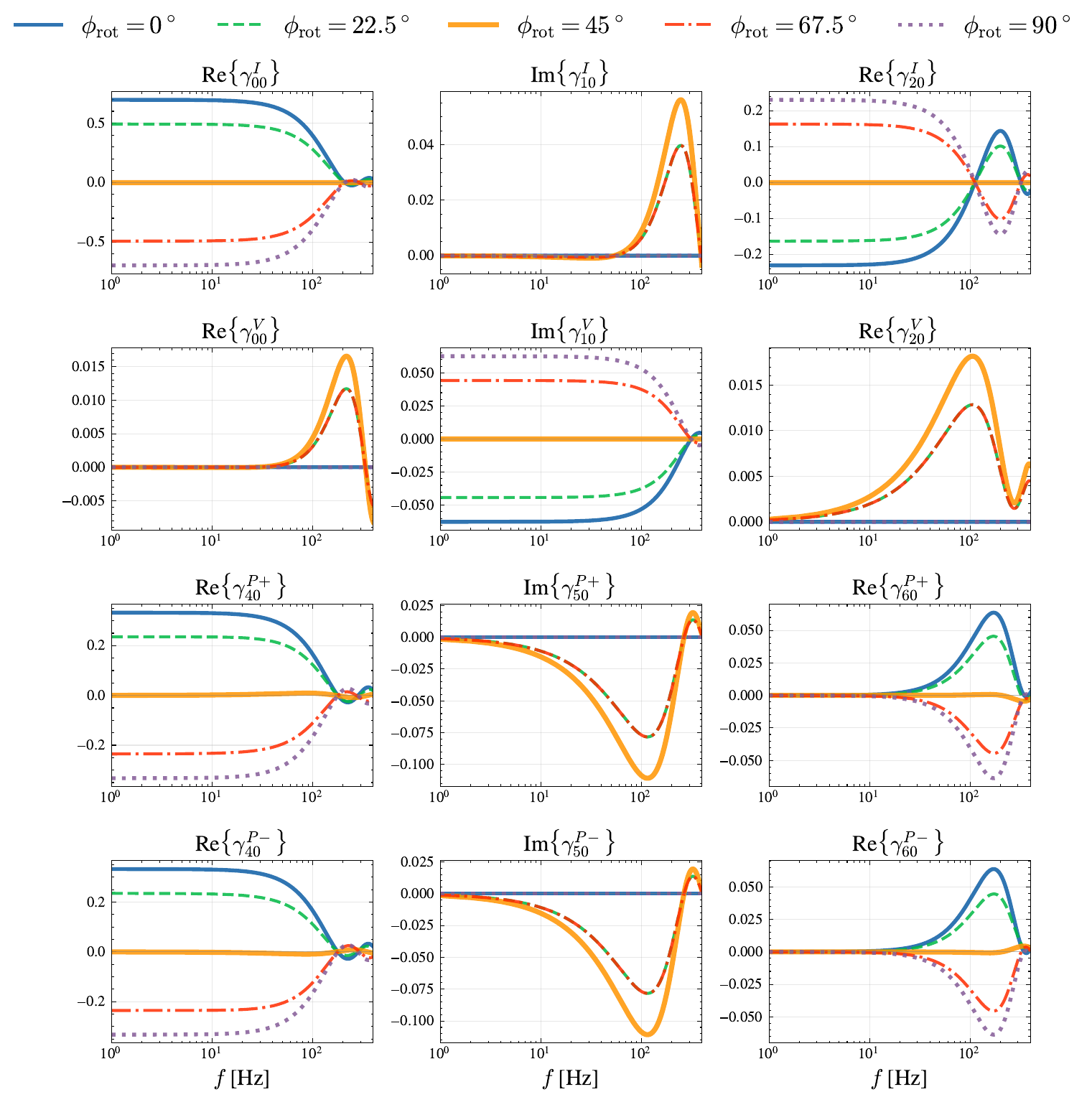}
  \caption{\black{Orientation dependence of the polarized ORFs for a planar
  two-detector geometry. The first two rows show $I$, the next two
  $V$, and the last two $P^+$ and $P^-$. Curves show
  $\gamma^X_{\ell 0}(f)$ for detector
  rotation angles $\phi_{\rm rot}=0$--$90^\circ$. Even-$\ell$ panels
  show the real part and odd-$\ell$ panels the imaginary part. The $I$
  and $P^\pm$ responses scale as $\cos 2\phi_{\rm rot}$ and vanish at
  $45^\circ$ while the even-$\ell$ $V$ response scales as
  $\sin 2\phi_{\rm rot}$ and is maximal there.}}
  \label{fig:orientation_orfs}
\end{figure*}

\black{Since these ORFs are functions of the relative orientation of the two detectors, we consider the setup of Ref.~\cite{Christensen:2026yyl}, two $L$-shaped detectors separated by $1000$ km on a sphere of Earth's radius, with the second rotated about its radial direction by $\Phi{\rm rot}$ from the aligned configuration. For $m=0$ we find that $\Gamma^{I,P+,P-}(f)$ vary as $\cos 2\Phi_{\rm rot}$ for even $\ell$ and as $\sin 2\Phi_{\rm rot}$ for odd $\ell$, and the $\Gamma^{V}(f)$'s the other way around, as shown in Fig.~\ref{fig:orientation_orfs}. Also, for even (odd) $\ell$ and $m=0$ the ORFs are purely real (imaginary). For $m\neq0$, there is no such simplification due to the arbitrary choice of sidereal phase.}

\section{Simulation Tests} \label{sec:inj}

\subsection{Map-making Validation}\label{sec:map-making}
We simulate a scenario with stationary Gaussian noise at an expected O5 sensitivity \cite{ALIGO} using the LIGO Hanford, LIGO Livingston, and Virgo network. The analysis is performed in both the spherical-harmonic and pixel bases with $\ell_{\max}=24$ and HEALPix resolution \cite{Gorski:2004by} $N_{\rm side}=8$ ($N_{\rm pix}=768$), and three years of data is folded \cite{Ain:2015lea} into one sidereal day using $\Gamma^{X}_{\I\J;\ell m}(f;t)=e^{i m\Omega_{\oplus}(t-t_0)}\Gamma^{X}_{\I\J;\ell m}(f;t_0)$ ($\Omega_{\oplus}$ is the Earth's angular rotational velocity). Maps are reported in units $\Omega_{X}(f,\nhat)\equiv \frac{2\pi^2 f^3}{3H_0^2}\,X(f,\nhat)$, with polarization maps parameterized by $(p,\chi,\varphi)$ as
\begin{align}
\Omega_Q&=\;\Omega_I\,p\sin\chi\,\cos(4\varphi),\nonumber\\
\Omega_U&=\;\Omega_I\,p\sin\chi\,\sin(4\varphi),\nonumber\\
\Omega_V&=\;\Omega_I\,p\cos\chi.
\end{align}

\textcolor{black}{Here $p\in[0,1]$ is the fraction of the intensity that is polarized, $\chi$ sets how this polarized power is shared between circular and linear polarization, so that $\chi=0$ gives purely circular and $\chi=90^\circ$ purely linear polarization, and $\varphi$ is the orientation angle of the linear polarization, which enters as $4\varphi$ because $P^\pm$ carry spin weight $\pm4$.}

The simulation consists of an unpolarized galactic plane (band half-width $10^\circ$
about the great circle with pole at $(\theta,\phi)=(45^\circ,0^\circ)$, where $\theta$ is colatitude) with
$\Omega_I(f)=10^{-6}(f/25\,{\rm Hz})^{2/3}$, a circularly polarized cap
centered at $(\theta,\phi)=(90^\circ,90^\circ)$ of radius $12^\circ$ with
$(p,\chi,\varphi)=(0.8,0^\circ,0^\circ)$, and a linearly polarized cap centered at
$(\theta,\phi)=(90^\circ,270^\circ)$ of radius $12^\circ$ with
$(p,\chi,\varphi)=(0.8,90^\circ,40^\circ)$.

\begin{figure*}[htbp]
  \centering
  \includegraphics[width=\textwidth]{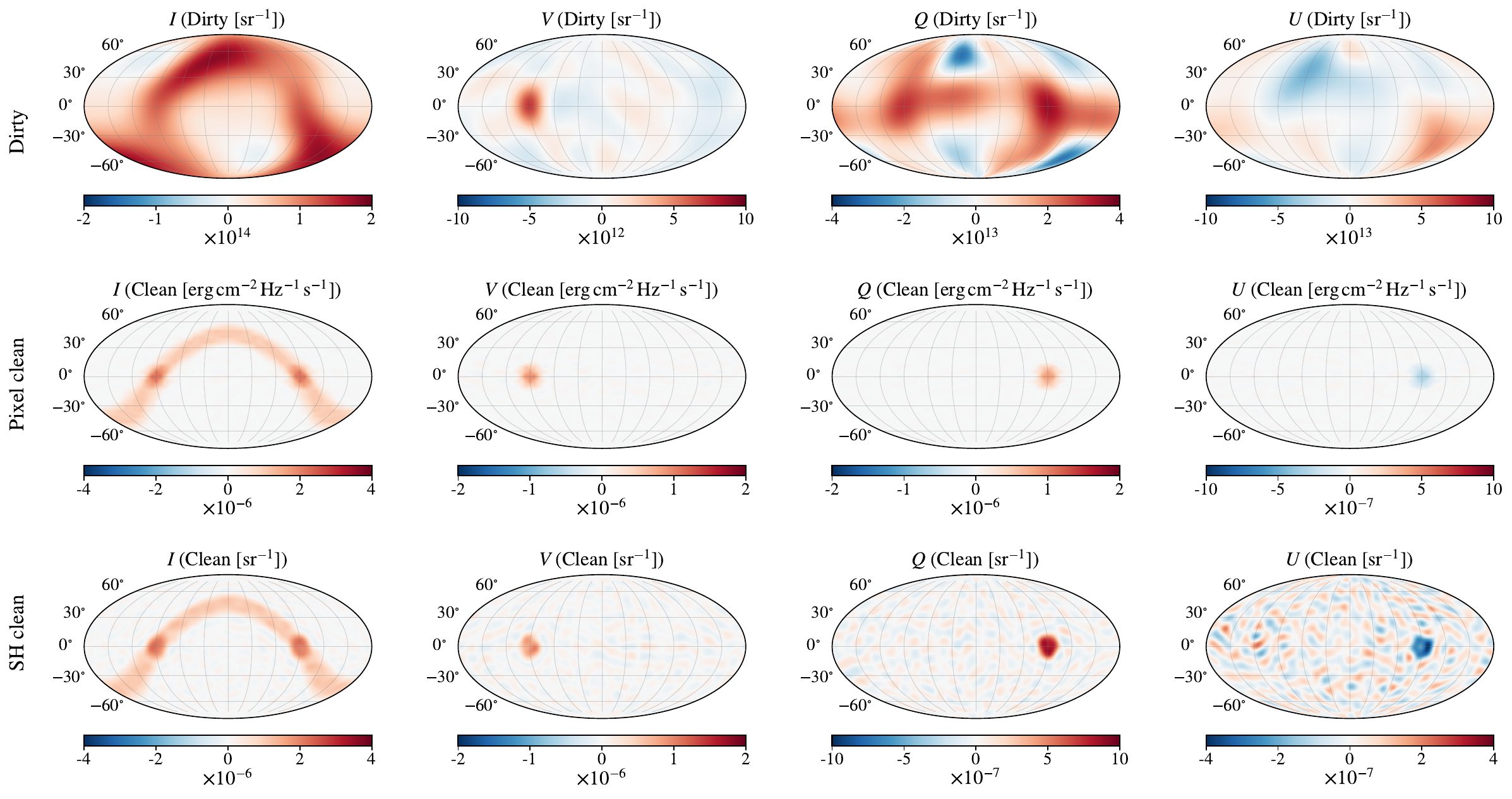}
  \caption{
  Benchmark full-Stokes map making for the simulated polarized background using the HLV network (three years; $\ell_{\max}=24$; $N_{\rm side}=8$). Columns show the reconstructed Stokes components $I$, $V$, $Q$, and $U$ in a Mollweide projection (coordinates are colatitude $\theta$ and longitude $\phi$, with latitude $b=90^\circ-\theta$). The three rows compare the network dirty maps $\hat{\mathcal{X}}$ to the corresponding clean maps in the pixel and spherical-harmonic bases; both bases recover the simulated morphology: an unpolarized band (half-width $10^\circ$) about the great circle with pole at $(\theta,\phi)=(45^\circ,0^\circ)$, a circularly polarized cap in $V$ centered at $(90^\circ,90^\circ)$ with radius $12^\circ$ and $(p,\chi,\varphi)=(0.8,0^\circ,0^\circ)$, and a linearly polarized cap in $Q/U$ centered at $(90^\circ,270^\circ)$ with radius $12^\circ$ and $(p,\chi,\varphi)=(0.8,90^\circ,40^\circ)$ (giving a stronger $Q$ feature than $U$).
  }
  \label{fig:maps}
\end{figure*}

\begin{figure*}[htbp]
    \centering
    \includegraphics[width=\linewidth]{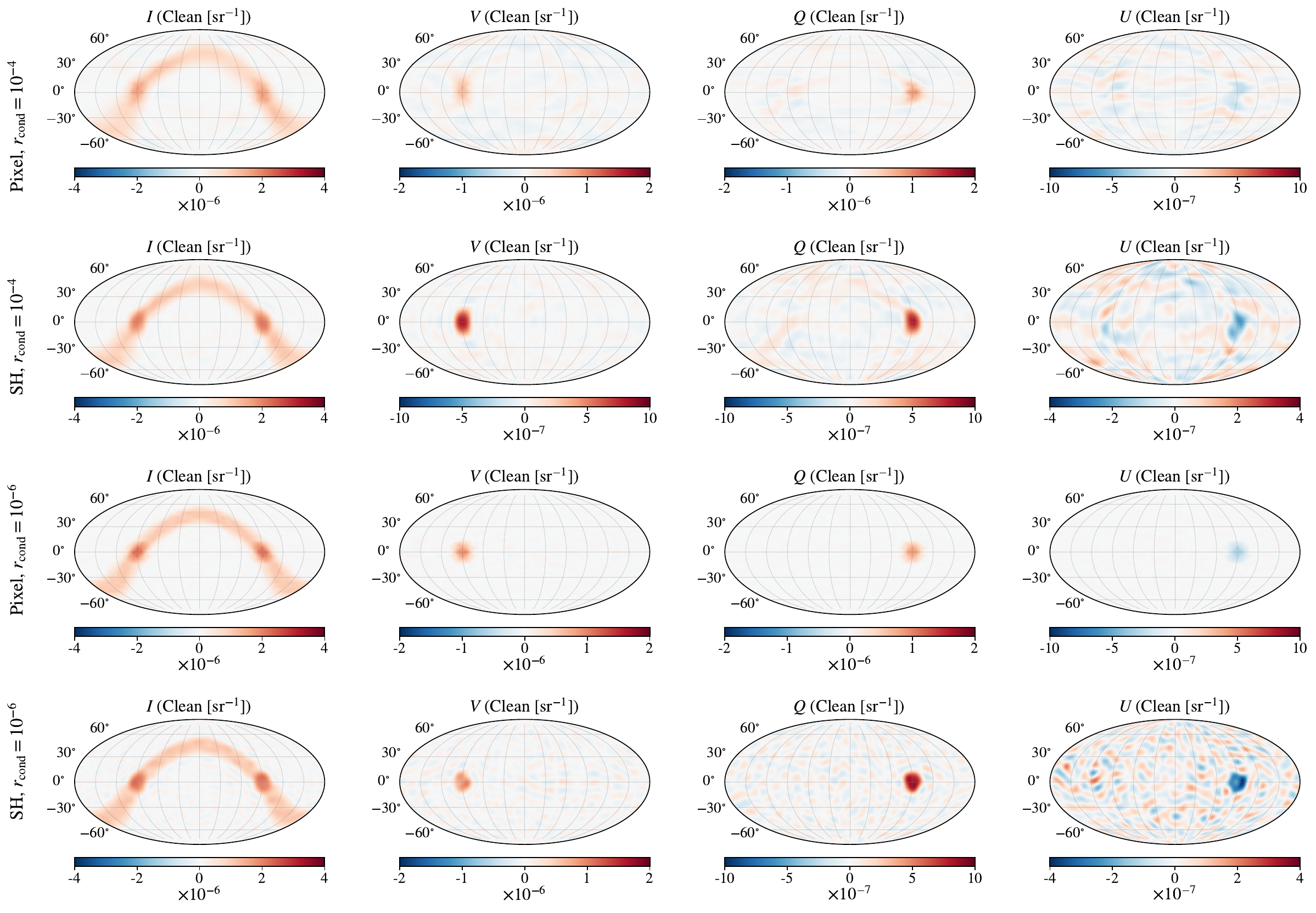}
    \caption{\black{Same as Fig.~\ref{fig:maps} using the eigenvalue-threshold cutoff $r_{\rm cond}=10^{-4}$ (upper half) and $r_{\rm cond}=10^{-6}$ (lower half). The dirty maps are independent of $r_{\rm cond}$ and are omitted.}}
    \label{fig:rcond_maps}
\end{figure*}

In this realization (Fig.~\ref{fig:maps}), the full-Stokes reconstruction recovers the simulated unpolarized plane in $I$ and the polarized caps in $V$ and $Q/U$ in both the pixel and spherical-harmonic implementations. Using the standard map-making estimator $\hat a=\mathcal{F}^{-1}\hat{\mathcal{X}}$, we define the coefficient covariance
$\boldsymbol{\Sigma}\equiv{\rm Cov}(\hat a)=\mathcal{F}^{-1}_{\rm reg}$,
where $\mathcal{F}^{-1}_{\rm reg}$ is the eigenvalue-threshold pseudo-inverse of $\mathcal{F}=U\,{\rm diag}(\lambda_i)\,U^\dagger$, retaining only modes with $\lambda_i/\lambda_{\max}\ge r_{\rm cond}$. We adopt $r_{\rm cond}=10^{-5}$ as a representative cutoff that removes poorly constrained modes while retaining the large-scale features of interest; moderate variations in $r_{\rm cond}$ ($10^{-4}$--$10^{-6}$) \black{as shown in Fig. \ref{fig:rcond_maps}} leave the qualitative structure unchanged, affecting only the expected variance--resolution tradeoff.

\begin{figure*}[htbp]
  \centering
  \includegraphics[width=\textwidth]{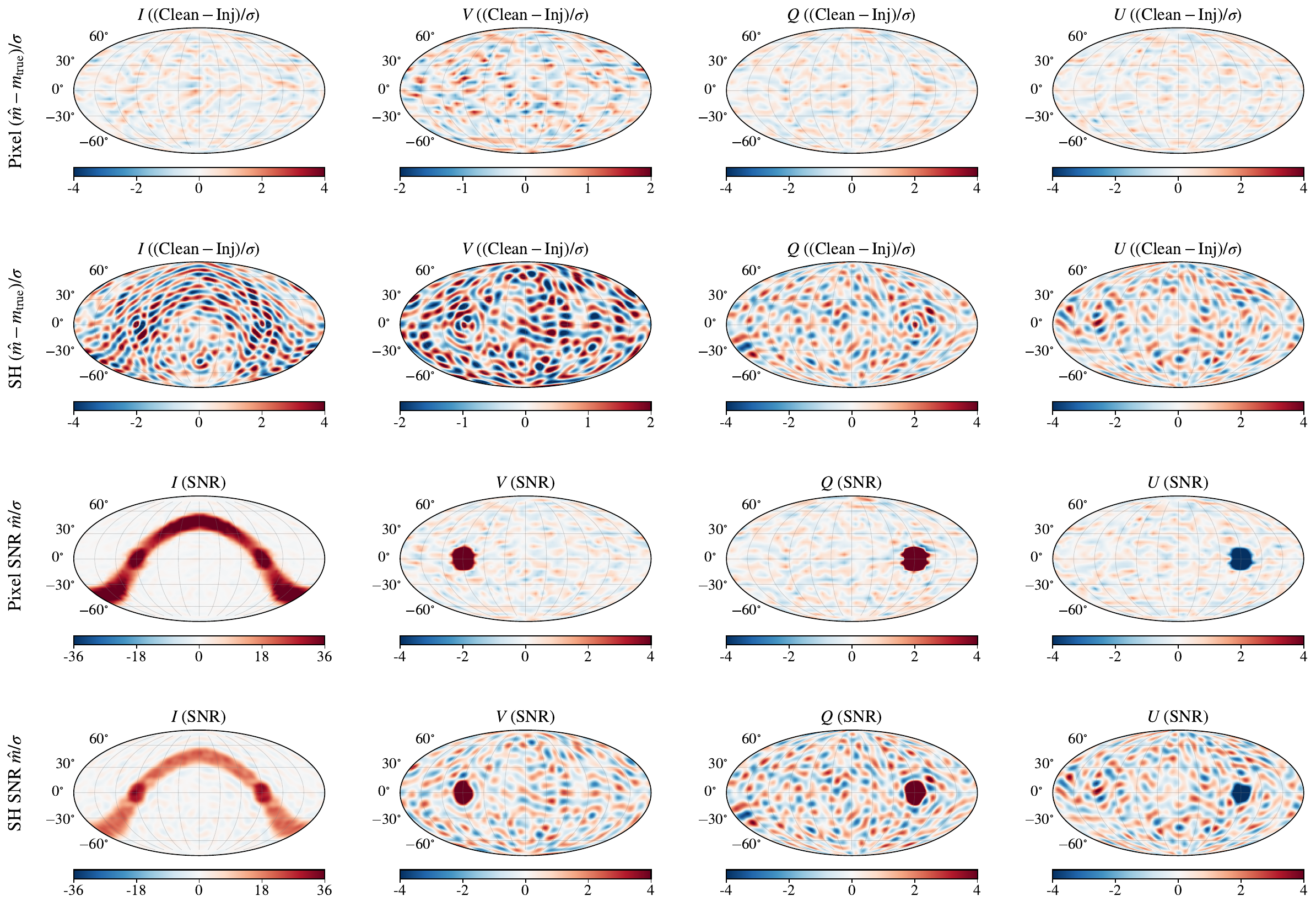}
  \caption{
  Benchmark full-Stokes map making (continued). The rows quantify reconstruction accuracy using the per-pixel uncertainty $\sigma^X_p\equiv\sqrt{\big(\boldsymbol{\Sigma}^{(m)}\big)^{XX}_{pp}}$ obtained by propagating the regularized coefficient covariance $\boldsymbol{\Sigma}=\mathcal{F}^{-1}_{\rm reg}$. We show normalized residuals $(\hat m^X_p-m^X_{{\rm true},p})/\sigma^X_p$ and the corresponding SNR maps ${\rm SNR}^X_p\equiv \hat m^X_p/\sigma^X_p$ for each basis. The residual-over-$\sigma$ maps are consistent with noise for the full-Stokes solution, while the SNR maps highlight the high-significance recovery of the plane in $I$ and the polarized caps (see the text).
  }
  \label{fig:maps:cont}
\end{figure*}

\begin{figure*}[htbp]
  \centering
  \includegraphics[width=\textwidth]{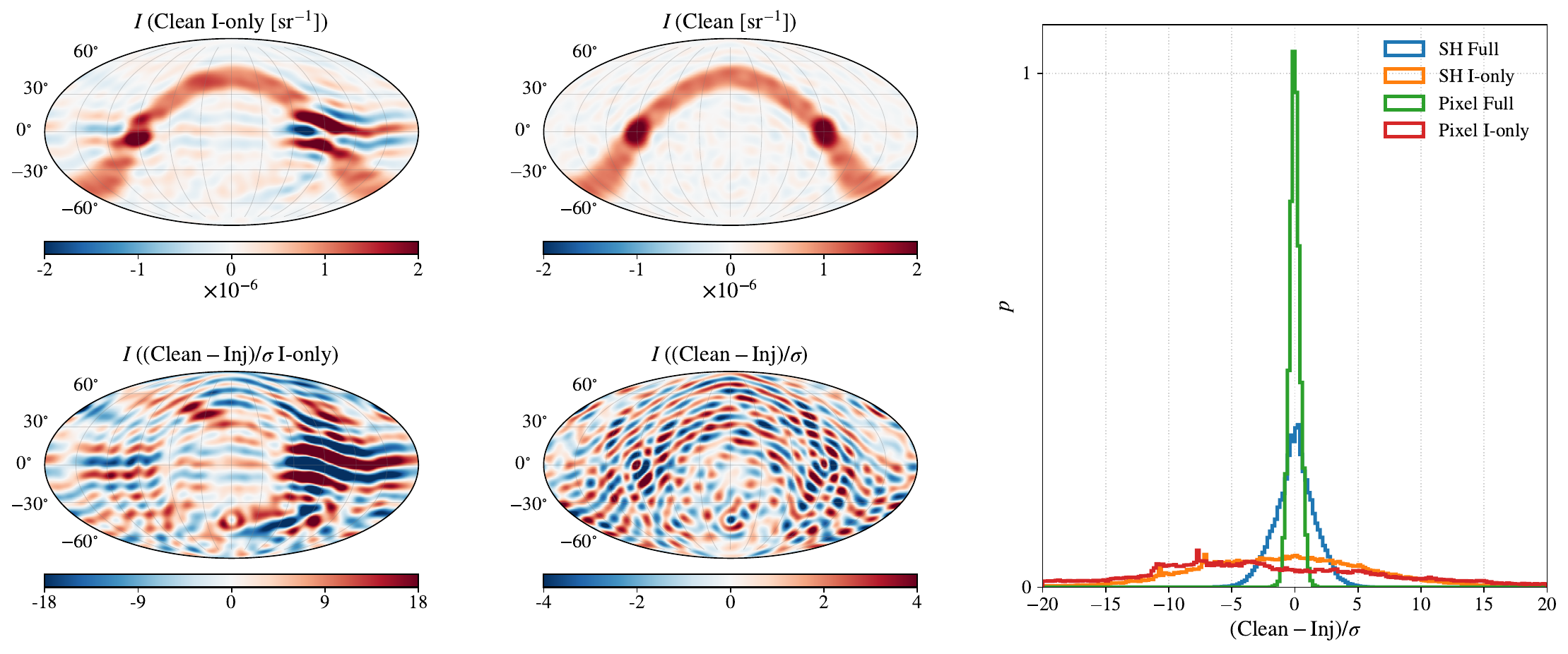}
  \caption{$I$-only versus full-Stokes reconstruction for the benchmark scenario. \emph{Left column:} $I$-only analysis; \emph{center column:} full-Stokes analysis. Top row shows the clean intensity map $\hat m^{I}$, bottom row shows the normalized residual $(\hat m^{I}-m^{I}_{\rm true})/\sigma^{I}$ (note the different color scales: $\pm 18$ for $I$-only vs.\ $\pm 4$ for full-Stokes). \emph{Right:} histogram of the per-pixel normalized residual for the $I$ Stokes parameter across both basis representations and analysis types.}
  \label{fig:bias_maps}
\end{figure*}

We denote the reconstructed pixelized Stokes maps by $\hat m^X_p$ ($X\in\{I,V,Q,U\}$), \black{with $\hat m^X_p\equiv \hat a^X_p$ in the pixel basis and $\hat m^X_p=\sum_{\ell m}\hat a^X_{\ell m}\,{}_{s_X}Y_{\ell m}(\hat n_p)$
evaluated on the same grid in the spherical-harmonic basis.} Propagating $\boldsymbol{\Sigma}$ to map space gives $\boldsymbol{\Sigma}^{(m)}\equiv{\rm Cov}(\hat m)$ and the per-pixel $1\sigma$ uncertainty map
$\sigma^X_p\equiv\sqrt{\big(\boldsymbol{\Sigma}^{(m)}\big)^{XX}_{pp}}$,
from which we form the normalized residual and SNR maps,
$(\hat m^X_p-m^X_{{\rm true},p})/\sigma^X_p$ and ${\rm SNR}^X_p\equiv \hat m^X_p/\sigma^X_p$.
As shown in Fig.~\ref{fig:maps} and Fig.~\ref{fig:maps:cont}, the plane reaches per-pixel ${\rm SNR}\sim\mathcal{O}(10)$ while the polarized caps reach ${\rm SNR}\sim\mathcal{O}(1$\text{--}$10)$, with $Q$ most significant (and $U$ lower, consistent with the simulated $\varphi=40^\circ$).  The SH maps appear somewhat grainier than the pixel versions, most noticeably for $U$ which has the smallest simulated amplitude; this reflects the different sets of modes retained by the eigenvalue-threshold regularization in the two parameterizations. Repeating the analysis with the standard $I$-only pipeline yields residuals that deviate from noise and a clean $I$ map biased relative to the simulated intensity (Fig.~\ref{fig:bias_maps}); see Sec.~\ref{sec:inj_bias_maps} for representative map-level diagnostics.

\subsection{I-only Comparison} \label{sec:inj_bias_maps}
To illustrate the bias coming from the polarization blocks directly in map space, we compare the recovered intensity clean map $\hat m^{I}$ and the normalized residuals $(\hat m^{I}-m^{I}_{\rm true})/\sigma^{I}$ for the benchmark simulation described in Sec.~\ref{sec:map-making} using both the full-Stokes and $I$-only analyses (with $\sigma^{I}$ the per-pixel uncertainty propagated from the Fisher-matrix covariance).  In the full-Stokes reconstruction the residual-over-$\sigma$ map is consistent with noise, with fluctuations centered near zero and typical tails at the $\sim2\sigma$ level.  In contrast, the $I$-only reconstruction shows coherent extended regions where the residual significantly exceeds the uncertainty, providing a map-level visualization of how neglected polarized anisotropies leak into the inferred intensity.

\subsection{Network Conditioning Tests}

In this section we demonstrate additional detectors added to the network improve the conditioning of the Fisher matrix.  We consider a stylized ``SGWB'' sky consisting of four
compact features labeled \{S,W,G,B\}: S is unpolarized (intensity only), W is
predominantly circularly polarized, and G/B are partially linearly polarized.
We analyze three years of folded data in the spherical-harmonic basis with
$\ell_{\max}=24$.

\begin{figure*}[ht]
    \centering
    \includegraphics[width=0.9\textwidth]{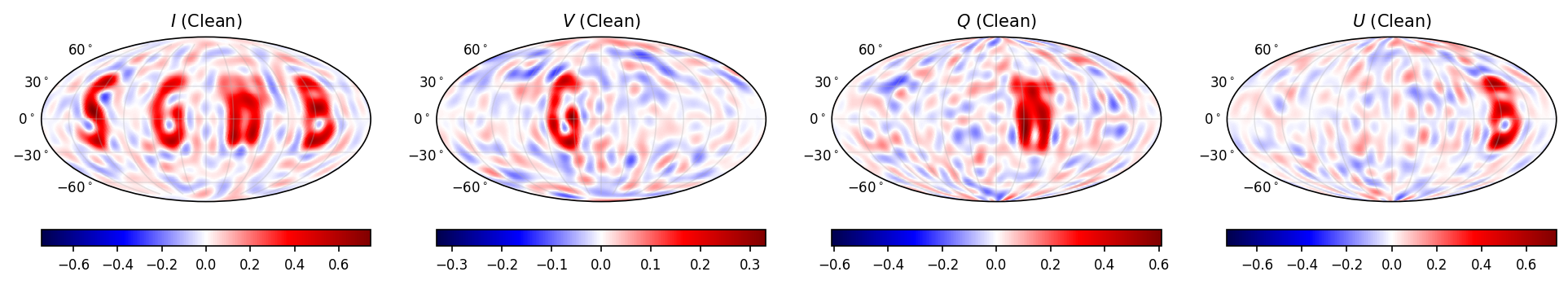}\\[2mm]
    \includegraphics[width=0.9\textwidth]{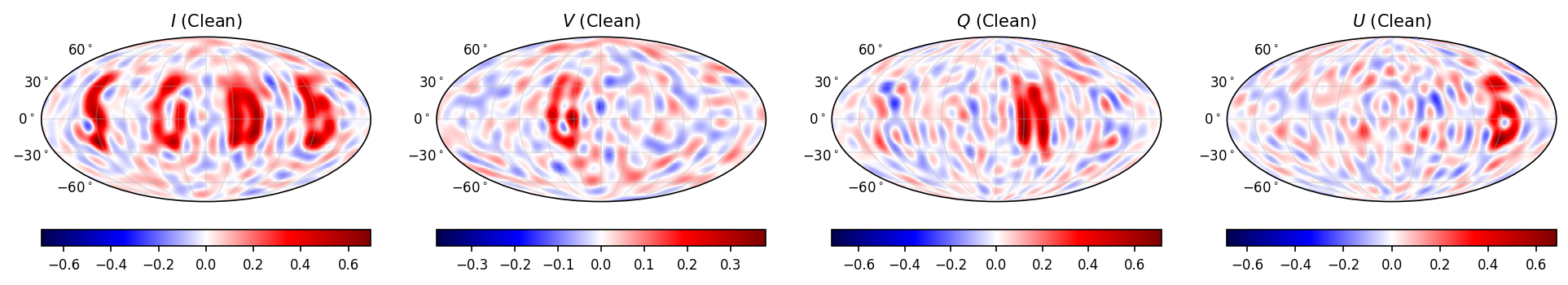}\\[2mm]
    \includegraphics[width=0.9\textwidth]{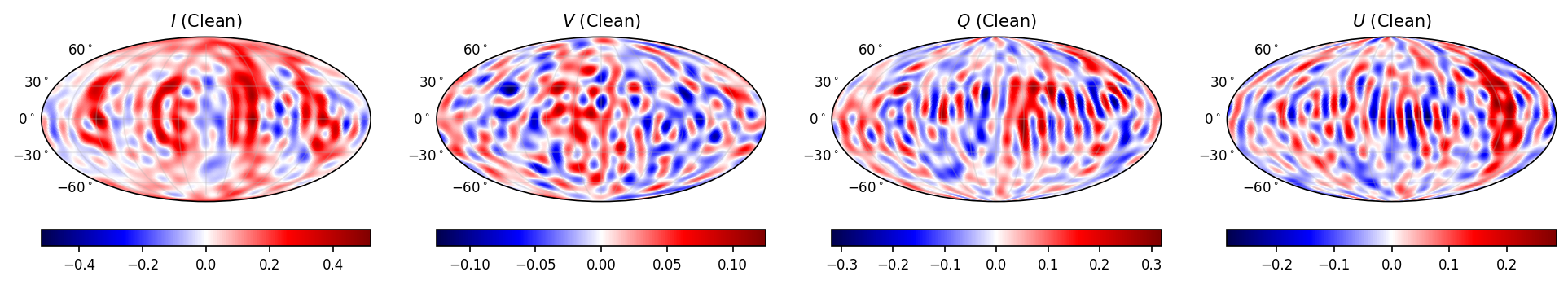}
    \caption{Example simulation in the spherical-harmonic basis for
    different detector networks (top to bottom: HLVK, HLV, HL) with
    $\ell_{\max}=24$ and $r_{\rm cond}=10^{-5}$. Color
    scales are chosen for visual clarity rather than absolute calibration.
    Fig.~\ref{fig:fisher_eigs} quantifies the corresponding conditioning via
    the Fisher-eigenvalue spectra.}
    \label{fig:SGWB1}
\end{figure*}

As in Sec.~\ref{sec:inj}, Fisher-matrix inversions are regularized via an
eigenvalue-threshold pseudo-inverse.  Writing
$\mathcal{F}=U\,\mathrm{diag}(\lambda_i)\,U^\dagger$, we retain only modes with
$\lambda_i/\lambda_{\max}\ge r_{\rm cond}$ and take $r_{\rm cond}=10^{-5}$ as a
representative benchmark.  We compare three detector networks with
decreasing baseline coverage: HLVK, HLV, and HL with expected O5 sensitivity \cite{ALIGO}.  The resulting recovered maps
are shown in Fig.~\ref{fig:SGWB1}; the qualitative recovery visibly degrades as
baselines are removed, consistent with increased degeneracy in the inversion of Fisher matrix.

\begin{figure}[t]
  \centering
  \includegraphics[width=0.8\linewidth]{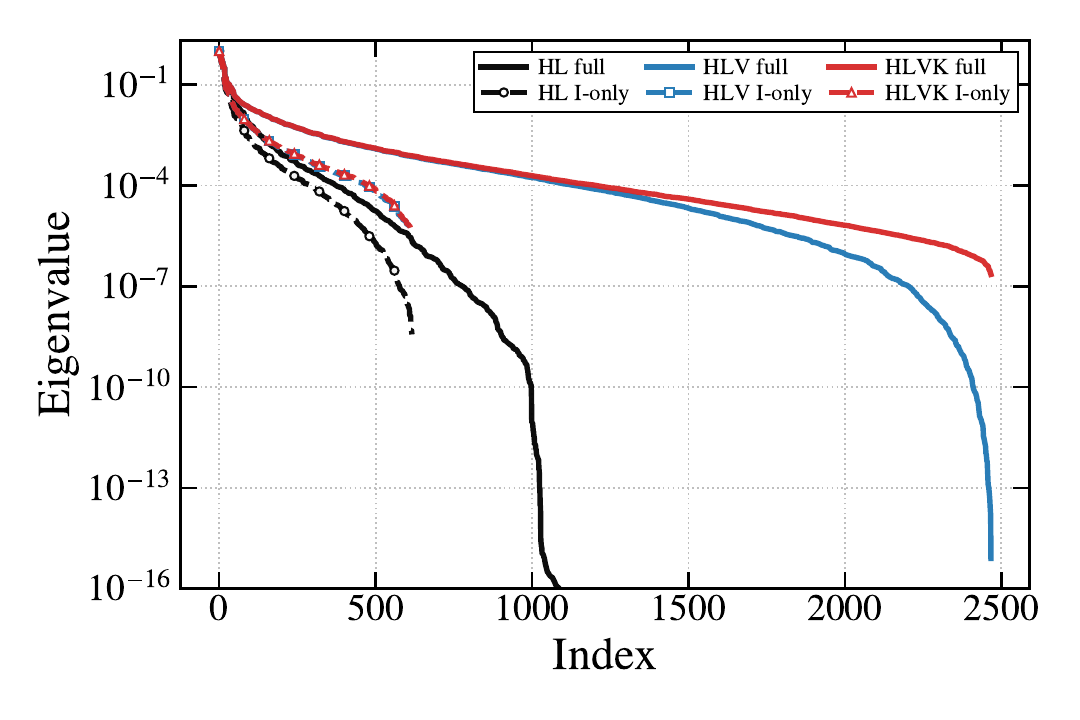}
  \caption{Sorted eigenvalues of the Fisher matrix for the
  different networks (HLVK, HLV, HL), each normalized by its maximum
  eigenvalue. Solid curves show the full-Stokes Fisher matrix; dashed curves
  show the intensity-only ($I$-block) Fisher matrix. Adding
  baselines increases the number of modes above threshold (improved
  conditioning/effective rank), with the strongest gains from the addition of Virgo.}
  \label{fig:fisher_eigs}
\end{figure}

To quantify this behavior independent of map amplitudes, Fig.~\ref{fig:fisher_eigs}
shows the sorted Fisher eigenvalue spectra for these cases, with
each spectrum normalized by its maximum eigenvalue.  Adding baselines increases
the number of modes above the regularization threshold and hence increases the
effective rank of the reconstruction.  We show both the full-Stokes Fisher
matrix and, for comparison, the intensity-only ($I$-block) Fisher matrix,
highlighting that the strongest conditioning gains from additional baselines
typically occur in the full-Stokes system.

\onecolumngrid
\section{Intensity covariance in the full-Stokes analysis} \label{sec:cov}

Here we derive the result quoted in the main text: for an unregularized
full-rank fit, the intensity covariance in the full Stokes fit is larger
than or equal to that of the intensity-only fit.

We split the clean–map parameters as
$\mathbf a=(\mathbf a_I,\mathbf a_Y)$ with $Y\in\{V,+,-\}$ and write the
Fisher matrix and dirty map in block form
\begin{align}
\mathcal F
&=
\begin{pmatrix}
\mathcal F_{II} & \mathcal F_{IY}\\[2pt]
\mathcal F_{YI} & \mathcal F_{YY}
\end{pmatrix},
&
\mathcal X
&=
\begin{pmatrix}
\mathcal X_I\\[2pt]
\mathcal X_Y
\end{pmatrix}.
\end{align}
The standard anisotropic analysis keeps only the intensity block,
\begin{align}
\hat{\mathbf a}^{\,(I\text{-only})}_I
  &= \mathcal F_{II}^{-1}\,\mathcal X_I,
&
\Sigma^{II,(I\text{-only})}
  &= \mathcal F_{II}^{-1}.
\end{align}
In the full analysis the intensity block of the inverse Fisher
matrix is given by the Schur complement
\begin{align}
\mathbb{S}
&\equiv
\mathcal F_{II}
- \mathcal F_{IY}\,\mathcal F_{YY}^{-1}\,\mathcal F_{YI},
\end{align}
so that
\begin{align}
\hat{\mathbf a}^{\,(\rm full)}_I
&=
\mathbb{S}^{-1}\!\left(
  \mathcal X_I
  - \mathcal F_{IY}\,\mathcal F_{YY}^{-1}\,\mathcal X_Y
\right), &
\Sigma^{II,(\rm full)}
&= \mathbb{S}^{-1}.
\end{align}
The term $\mathcal F_{IY}\,\mathcal F_{YY}^{-1}\,\mathcal X_Y$ subtracts the polarization contribution leaked into the intensity dirty map via the off-diagonal Fisher blocks.

To compare the covariances, we write
\begin{align}
\mathbb{S}
&= \mathcal F_{II}^{1/2}\,(\mathbb{I}-M)\,\mathcal F_{II}^{1/2},\\
M
&\equiv
\mathcal F_{II}^{-1/2}\,\mathcal F_{IY}\,\mathcal F_{YY}^{-1}\,
\mathcal F_{YI}\,\mathcal F_{II}^{-1/2}.
\end{align}
The matrix $M$ is Hermitian and positive semi–definite: defining
$B\equiv \mathcal F_{YY}^{-1/2}\,\mathcal F_{YI}\,\mathcal F_{II}^{-1/2}$
we can write $M=B^\dagger B$, so all its eigenvalues are real and
non–negative.  For an unregularized full-rank fit, $\mathcal F$ is positive definite on the fitted parameter space. The Schur complement $\mathbb{S}$ is then positive
definite, which implies that all eigenvalues of $M$ are strictly smaller than
one and $\mathbb{I}-M$ is positive definite.

The intensity covariance in the full fit is then
\begin{align}
\Sigma^{II,(\rm full)}
&= \mathbb{S}^{-1}
 = \mathcal F_{II}^{-1/2}\,(\mathbb{I}-M)^{-1}\,\mathcal F_{II}^{-1/2}\\
&= \mathcal F_{II}^{-1}
 + \mathcal F_{II}^{-1/2}\big[(\mathbb{I}-M)^{-1}-\mathbb{I}\big]\mathcal F_{II}^{-1/2}.
\end{align}
Diagonalizing $M$ as $M = U \Lambda U^\dagger$ with eigenvalues
$\lambda_i\in[0,1)$, we see that the eigenvalues of
$(\mathbb{I}-M)^{-1}-\mathbb{I}$ are
\begin{align}
\frac{1}{1-\lambda_i}-1 = \frac{\lambda_i}{1-\lambda_i} \ge 0,
\end{align}
so $(\mathbb{I}-M)^{-1}-\mathbb{I}$ is positive semi–definite.  Therefore the second term
in the last line is a positive semi–definite matrix, and we conclude that
$\Sigma^{II,(\rm full)} - \Sigma^{II,(I\text{-only})}$ is positive semi–definite as well.

Thus the full covariance is equal to the intensity-only covariance
plus an extra non–negative contribution.  Any linear combination of
intensity multipoles therefore has an uncertainty in the full fit
that is greater than or equal to the corresponding uncertainty in the
intensity-only fit, as stated in the main text (see \cite{Gair:2015hra} for the PTA context of this generic behavior).

\FloatBarrier
\section{O3 Supplementary Results} \label{sec:o3_extra}

\begin{figure*}[t]
    \centering
    \includegraphics[width=\linewidth]{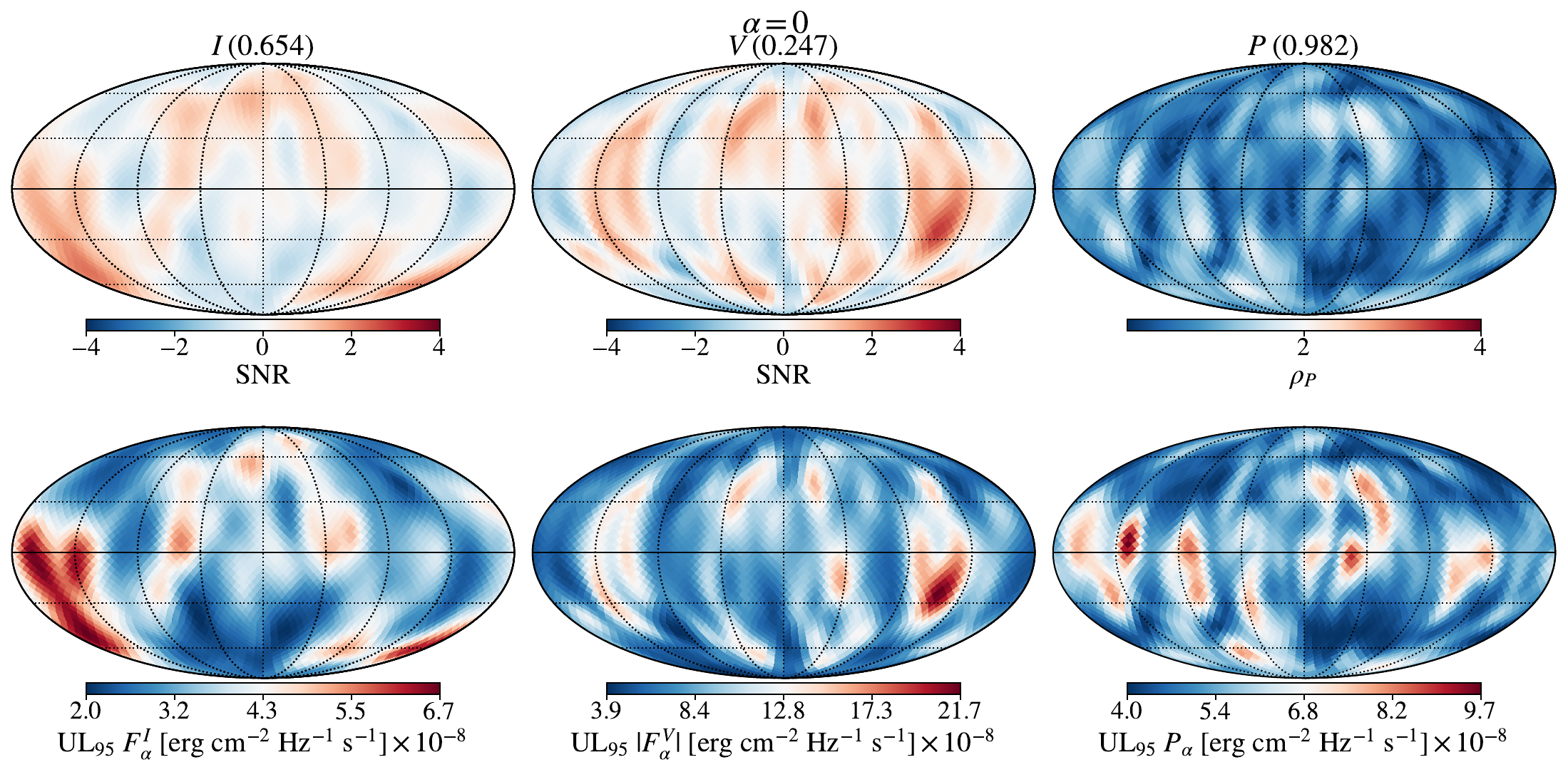}
    \includegraphics[width=\linewidth]{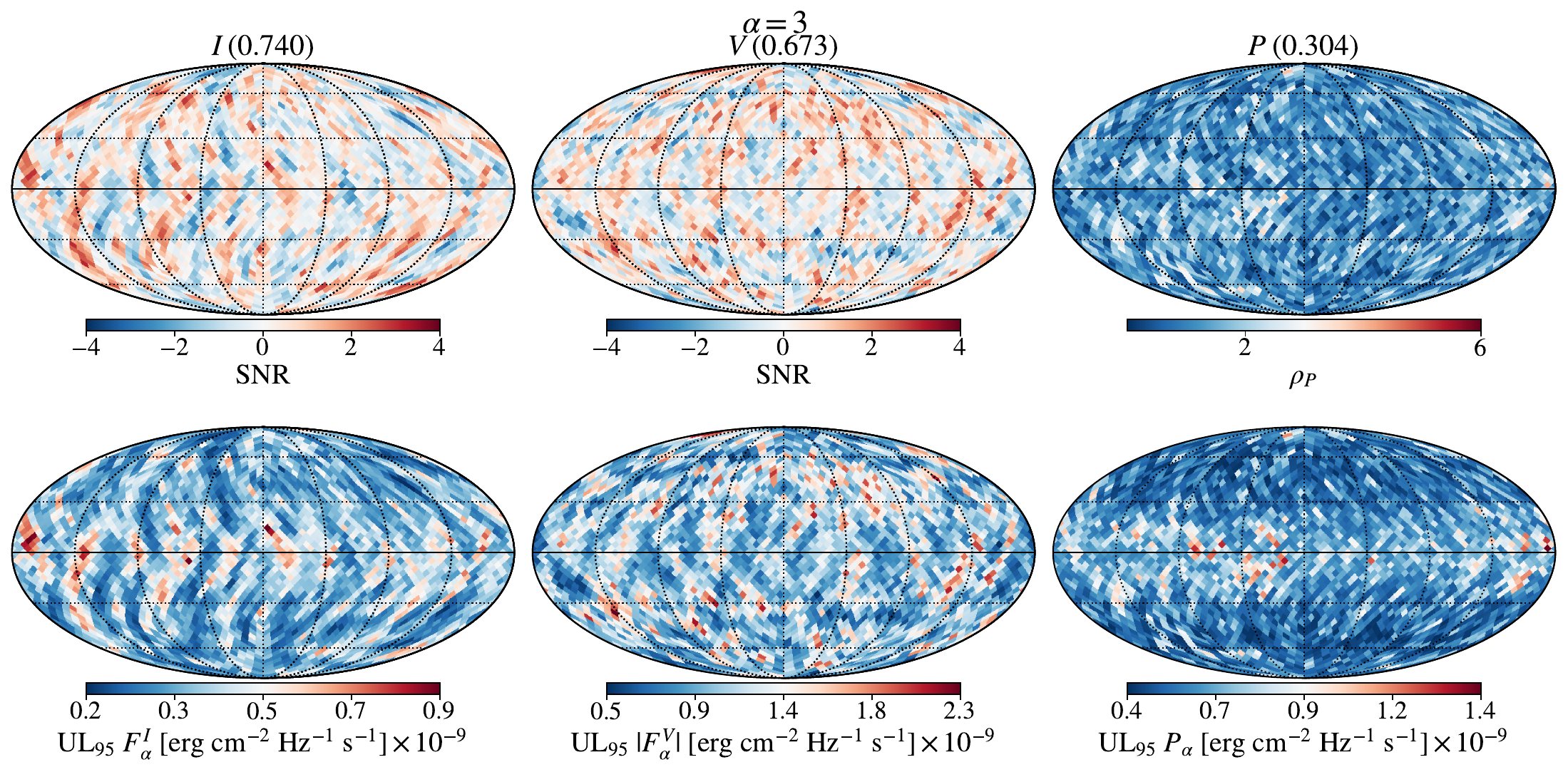}
    \caption{Full-Stokes O3 sky maps. The two upper rows correspond to $\alpha=0$ and the two lower rows to $\alpha=3$. In each pair the first row is the per-pixel SNR for $I$, $V$ and the linear-polarization significance $\rho_P$, and the second row is the corresponding $95\%$ upper limit on the GW energy-flux density $F^I_\alpha$, $|F^V_\alpha|$ and $P_\alpha=\sqrt{Q^2+U^2}$ in $\mathrm{erg\;cm^{-2}\;Hz^{-1}\;s^{-1}}$. Maps use $N_{\mathrm{side}}=16$; the significance and upper-limit construction is detailed in Sec.~\ref{sec:o3_methods}.}
    \label{fig:o3_extra}
\end{figure*}

When presenting O3 pixel-basis upper limits we report the GW energy flux density
\begin{equation}\label{eq:flux_def}
    F_\alpha(\nhat) = \frac{c\,\rho_c}{f_{\mathrm{ref}}}\,\Omega_\alpha(\nhat),
\end{equation}
where $\rho_c = 3H_0^2 c^2/(8\pi G)$ and $f_{\mathrm{ref}} = 25\;\mathrm{Hz}$. All other maps and $C_\ell$ results are reported as $\Omega$. Fig.~\ref{fig:o3_extra} presents the full-Stokes O3 sky maps for $\alpha=0$ and $\alpha=3$, complementing the $\alpha=2/3$ results of Fig.~\ref{fig:o3_wide_results}. We find no statistically significant anisotropy in any Stokes parameter at any spectral index. The polarized channels are therefore reported as upper limits throughout this section.

\subsection{Significance and upper-limit construction}\label{sec:o3_methods}

\begin{figure*}[htbp]
  \centering
  \includegraphics[width=0.96\linewidth]{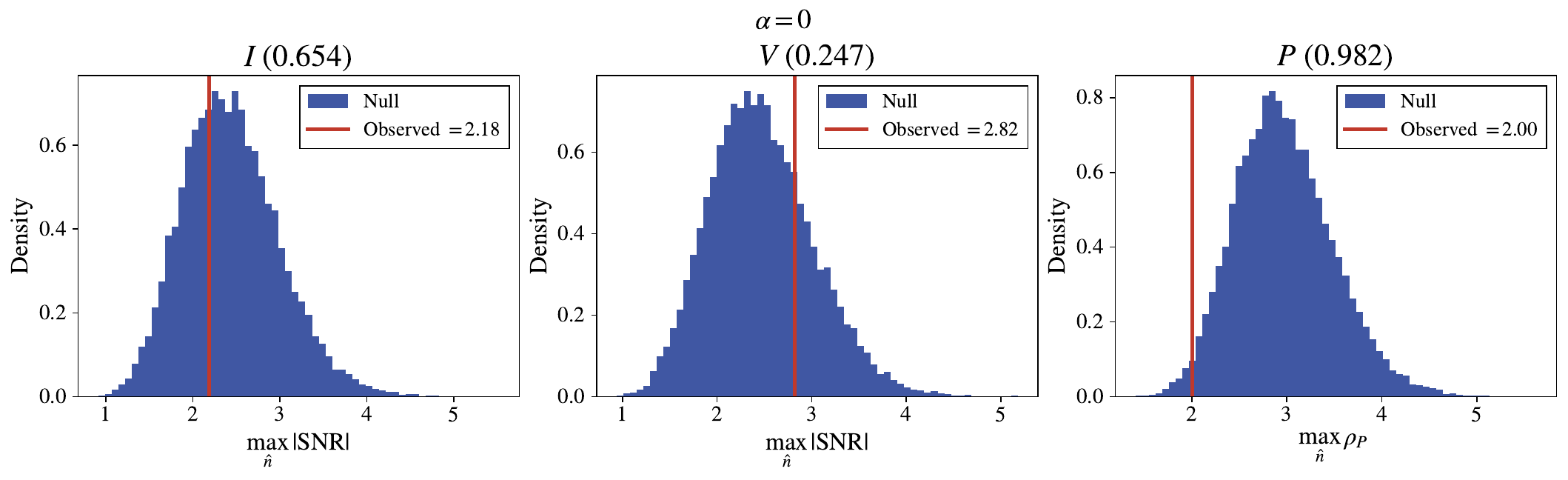}\\[1.5mm]
  \includegraphics[width=0.96\linewidth]{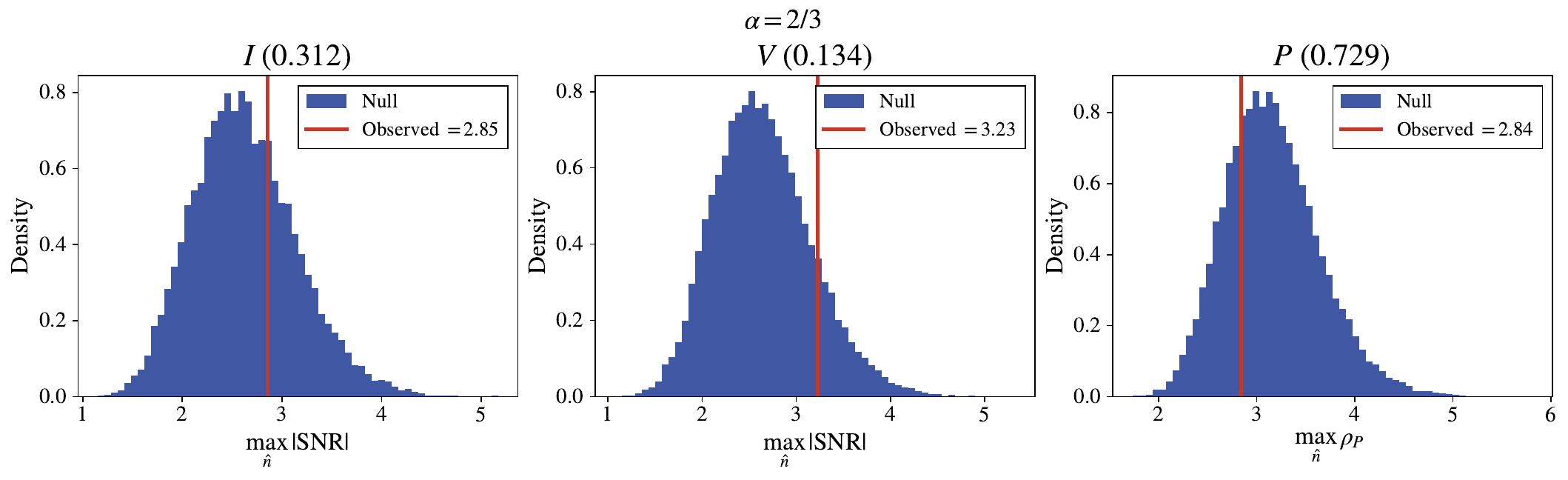}\\[1.5mm]
  \includegraphics[width=0.96\linewidth]{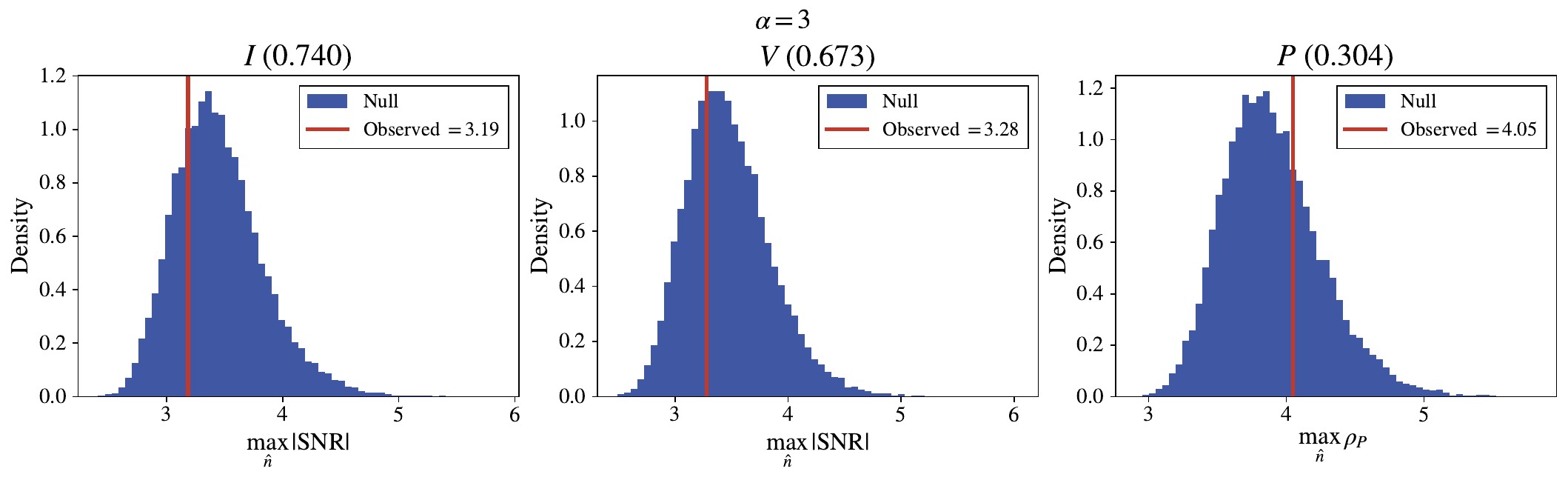}
    \caption{Pixel-basis global significance. From top to bottom the rows
    are $\alpha=0$, $2/3$ and $3$; each row shows the $I$, $V$ and $P$
    channels. The histogram is the noise-only null distribution of the
    sky maximum statistic ($\max_{\hat n}|\mathrm{SNR}|$ for $I,V$ and
    $\max_{\hat n}\rho_P$ for $P$) and the vertical line the observed
    value for our analysis. $p_{\mathrm{glob}}$ is the fraction of null realizations at or
    above this value.}
  \label{fig:o3_pix_pval}
\end{figure*}

The global significance of the per-pixel statistics defined in the main
text ($\mathrm{SNR}$ for $I,V$ and $\rho_P$ for linear polarization)
comes from a noise-only Monte Carlo with the full Fisher covariance: we
draw $X_{\mathrm{noise}}\sim\mathcal N(0,\Gamma)$ by colouring white
noise with the SVD of $\Gamma$, recompute the sky maximum
($T=\max_{\hat n}|\mathrm{SNR}|$ for $I,V$ and $T=\max_{\hat n}\rho_P$
for $P$) for each of $N_{\mathrm{MC}}=2\times10^{4}$ realizations, and
report $p_{\mathrm{glob}}=(1+\#\{T_{\mathrm{MC}}\ge T_{\mathrm{obs}}\})/(1+N_{\mathrm{MC}})$.
This null carries the sky-correlation look-elsewhere correction. The $95\%$ flux upper limits are flat non-negative Bayesian limits
with the per-detector amplitude calibration marginalized as in
Ref.~\cite{Whelan:2012zw}. Each detector $\I$ carries a calibration
factor $\lambda_\I\sim\mathcal N(1,\delta_\I^{2})$, with
$\delta_\I\in\{0.048,0.054\}$ (O1 H/L),
$\{0.026,0.0385\}$ (O2 H/L) and
$\{0.0696,0.0637,0.050\}$ (O3 H/L/V)~\cite{KAGRA:2021kbb,KAGRA:2021mth}.
Baseline factors $\lambda_{\I\J}=\lambda_\I\lambda_\J$ share detectors,
so we marginalize the joint per-baseline Gaussian likelihood for
$\hat F_{\I\J}$ over $\lambda_\I$ by Monte Carlo with
$N_\lambda=800$ draws, and report the $95\%$ quantile as the upper
limit. For $P=\sqrt{Q^2+U^2}$ the $Q$ and $U$ estimates enter jointly
with the polarization angle marginalized analytically. Figure~\ref{fig:o3_pix_pval} shows the resulting null
distributions and observed statistics.

\begin{figure*}[htbp]
    \centering
    \includegraphics[width=\linewidth]{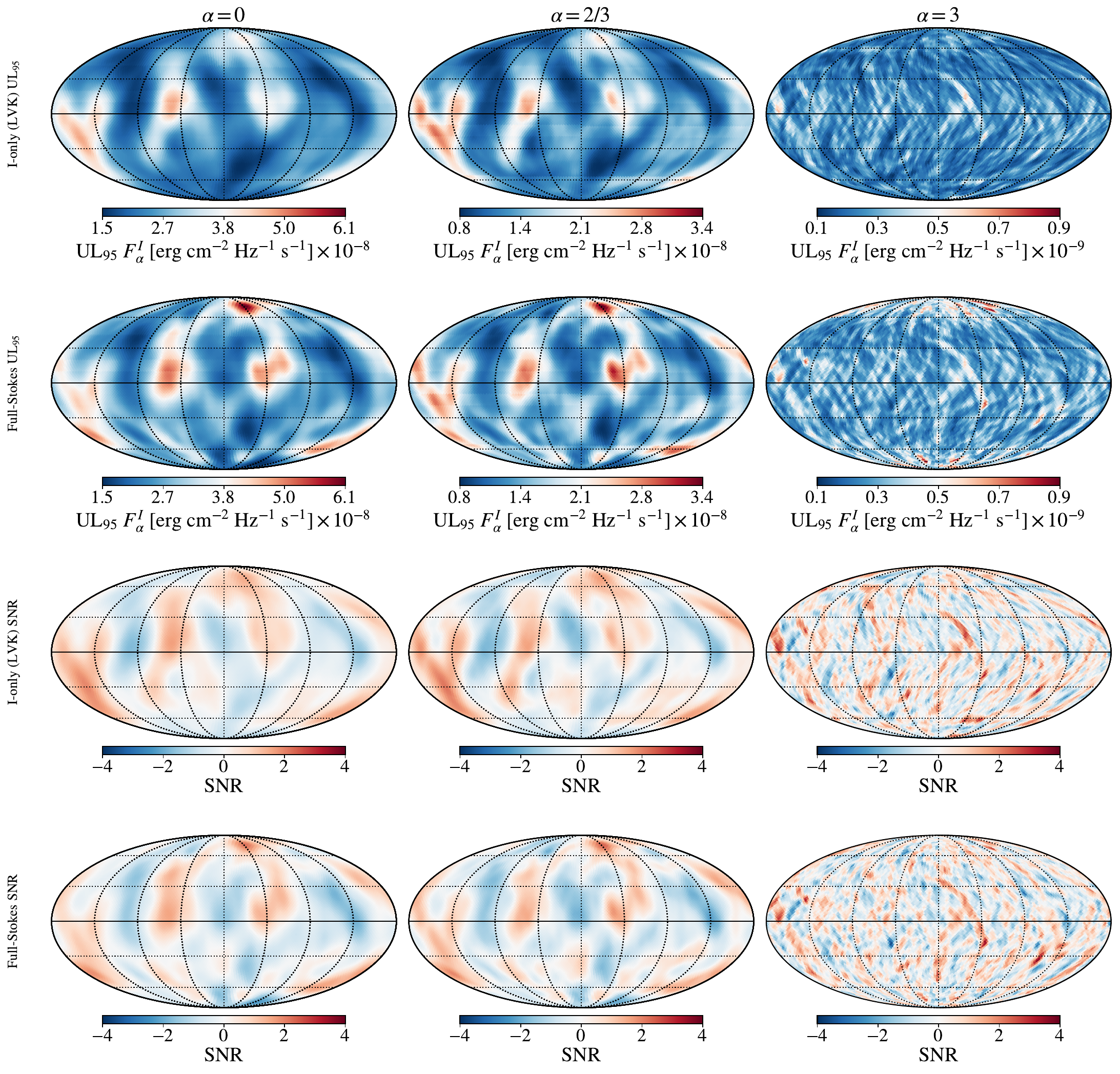}\\[1.5mm]
    \includegraphics[width=\linewidth]{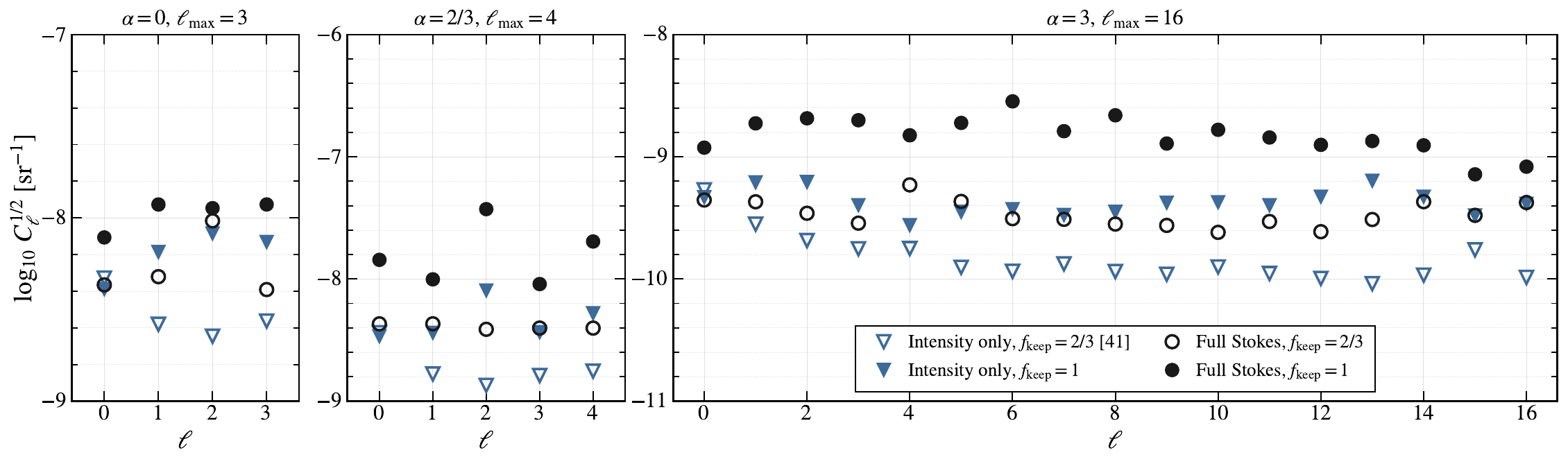}
    \caption{\black{Comparison of the $95\%$ upper limits and the SNR maps of the intensity-only analysis of Ref.~\cite{KAGRA:2021mth} and the polarized search of this paper. The first four rows show the broadband pixel-basis results at $\alpha=0,2/3,3$. The bottom row shows the spherical harmonics $C^{1/2}_\ell$ upper limits at each $\alpha$, where $f_{\rm keep}$ is the retained fraction of Fisher eigenmodes.}}
    \label{fig:lvk_vs_full}
\end{figure*}

The SPH analysis follows the LVK SPH construction of Refs.~\cite{KAGRA:2021mth,Suresh:2020khz,Ain:2018zvo} and inverts the joint Fisher matrix. \black{Its condition numbers are $3.9\times10^{2}$, $3.9\times10^{3}$ and $1.2\times10^{4}$ at $\ell_{\max}=3,4,16$ for $\alpha=0,2/3,3$, so no eigenmode is removed (we also show the regularized inversion comparison in Fig. \ref{fig:lvk_vs_full}),} with the angular-power estimator of Eq. \eqref{eq:cl_est} extended over Stokes indices $X,Y\in{I,V,E,B}$.

\begin{figure*}[htbp]
  \centering
  \includegraphics[width=\linewidth]{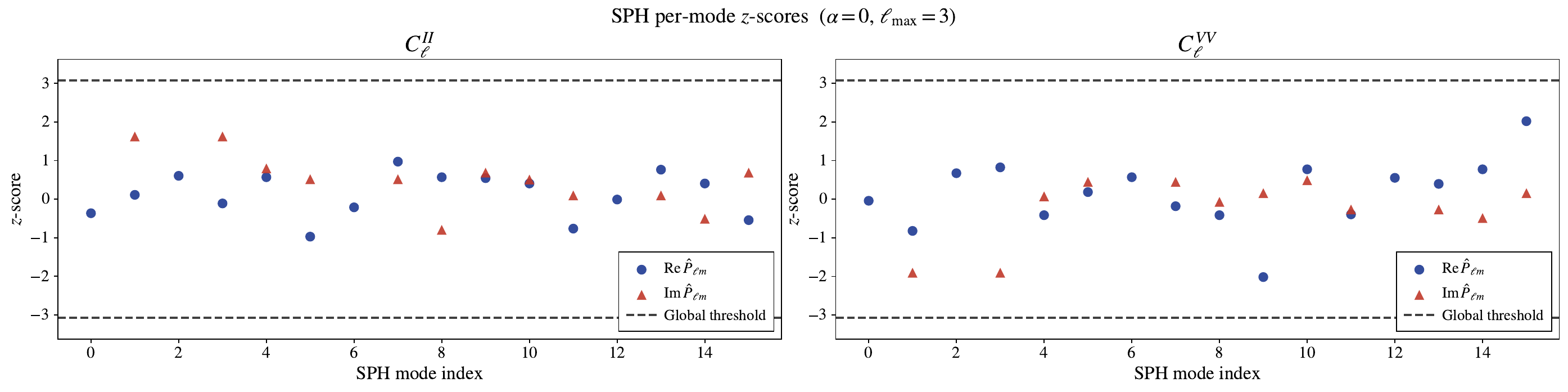}\\[1.5mm]
  \includegraphics[width=\linewidth]{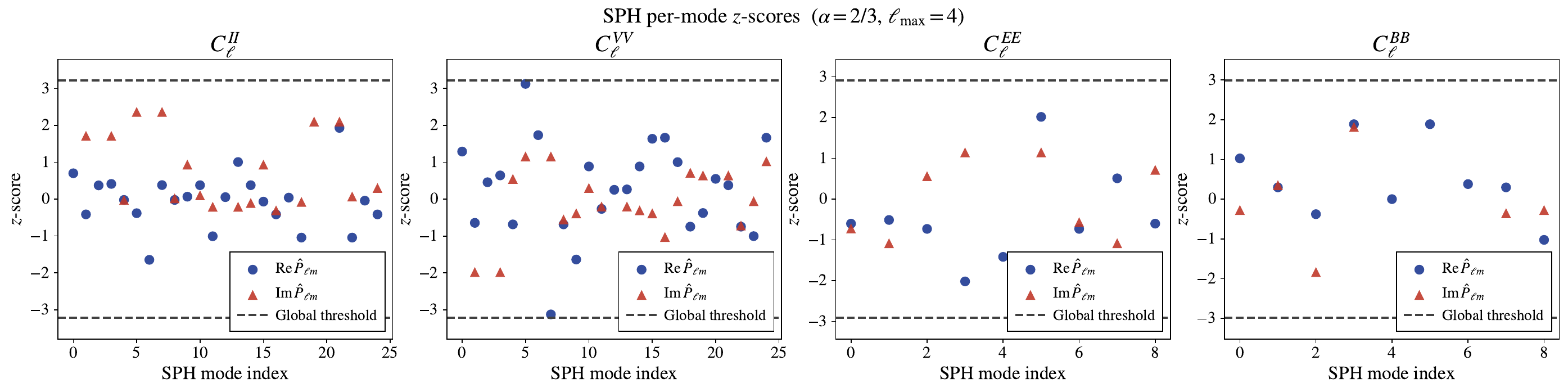}\\[1.5mm]
  \includegraphics[width=\linewidth]{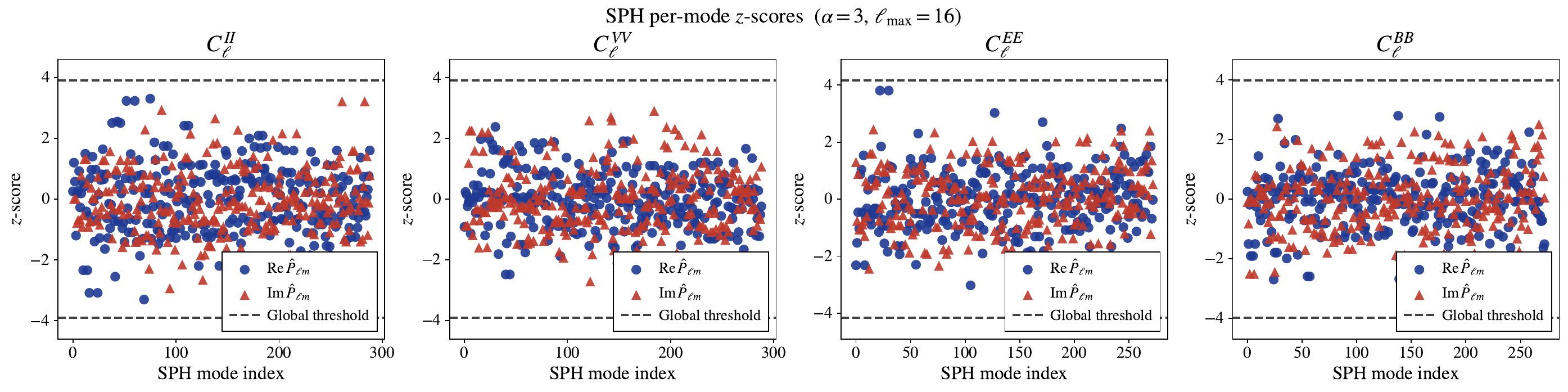}
    \caption{SPH per-mode $z$-scores; rows are $\alpha=0$, $2/3$ and $3$
    with $\ell_{\max}=3,4,16$ from top to bottom. Circles and triangles
    are the real and imaginary parts of $\hat\Omega_{\ell m}$; the dashed
    line is the $5\%$ quantile of the null $\max|z|$ distribution.}
  \label{fig:o3_sph_z}
\end{figure*}

For each Stokes spectrum we test whether any harmonic mode shows
significant deviation from noise. The clean-map coefficients
$\hat\Omega_{\ell m}$ are complex, with Re/Im variances
$\sigma^2_{{\rm Re},\ell m}$ and $\sigma^2_{{\rm Im},\ell m}$
following from the Fisher inverse~\cite{KAGRA:2021mth}. We form
\begin{equation}
  z^{\rm Re}_{\ell m}=\frac{\mathrm{Re}\,\hat\Omega_{\ell m}}{\sigma_{{\rm Re},\ell m}},
  \qquad
  z^{\rm Im}_{\ell m}=\frac{\mathrm{Im}\,\hat\Omega_{\ell m}}{\sigma_{{\rm Im},\ell m}},
\end{equation}
and take the largest $|z|$ across all modes as the test statistic.
These are standardized mode by mode, but the Fisher inversion
correlates them, so we obtain the distribution
of $\max|z|$ directly by Monte Carlo. We draw $5\times10^4$ dirty-map realizations
$\hat{\mathcal X}_{\rm sim}\sim\mathcal{CN}(0,\mathcal F)$, pass each
through the same Fisher inversion, and record the maximum $|z|$.
Comparing the observed maximum to this empirical null gives the
global $p$-value $p_{\rm glob}$. With a $5\%$ detection threshold we find
$p_{\rm glob}\in[0.07,0.93]$ across $\alpha\in\{0,2/3,3\}$ and Stokes
spectra, with no spectrum crossing threshold
(Fig.~\ref{fig:o3_sph_z}).

\begin{figure*}[htbp]
  \centering
  \includegraphics[width=\linewidth]{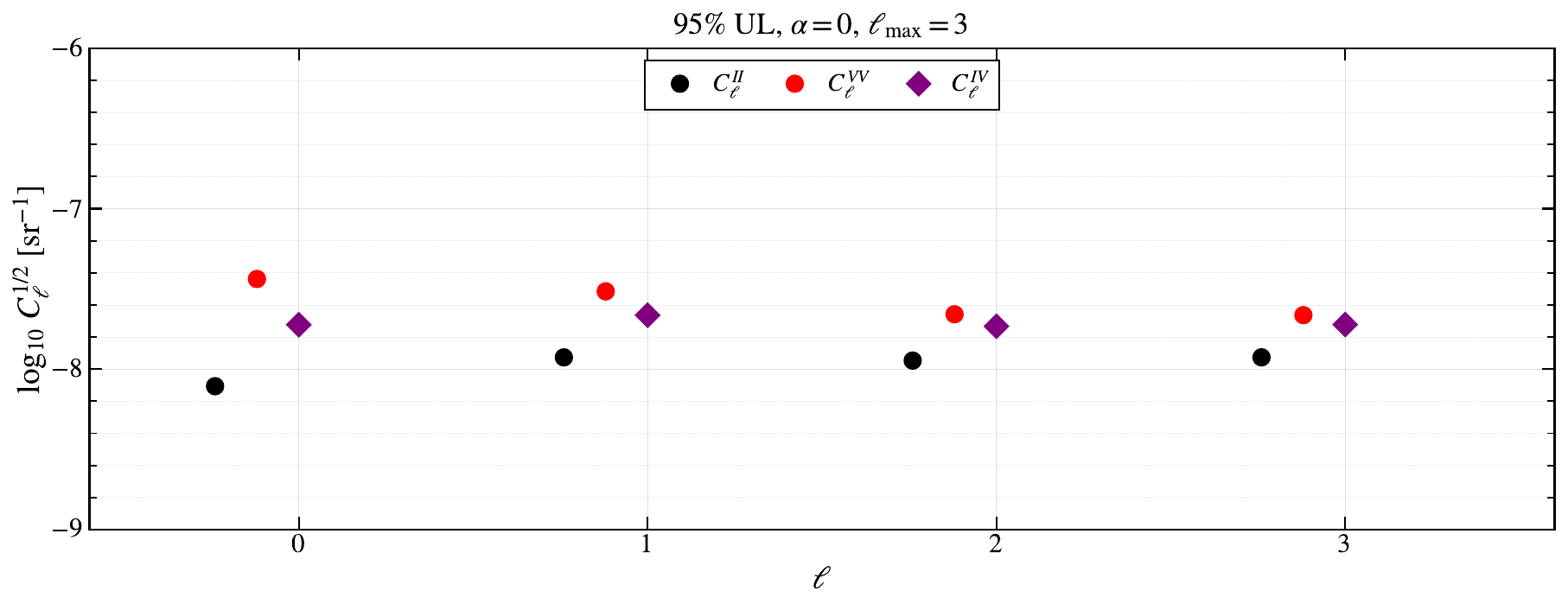}\\[1.5mm]
  \includegraphics[width=\linewidth]{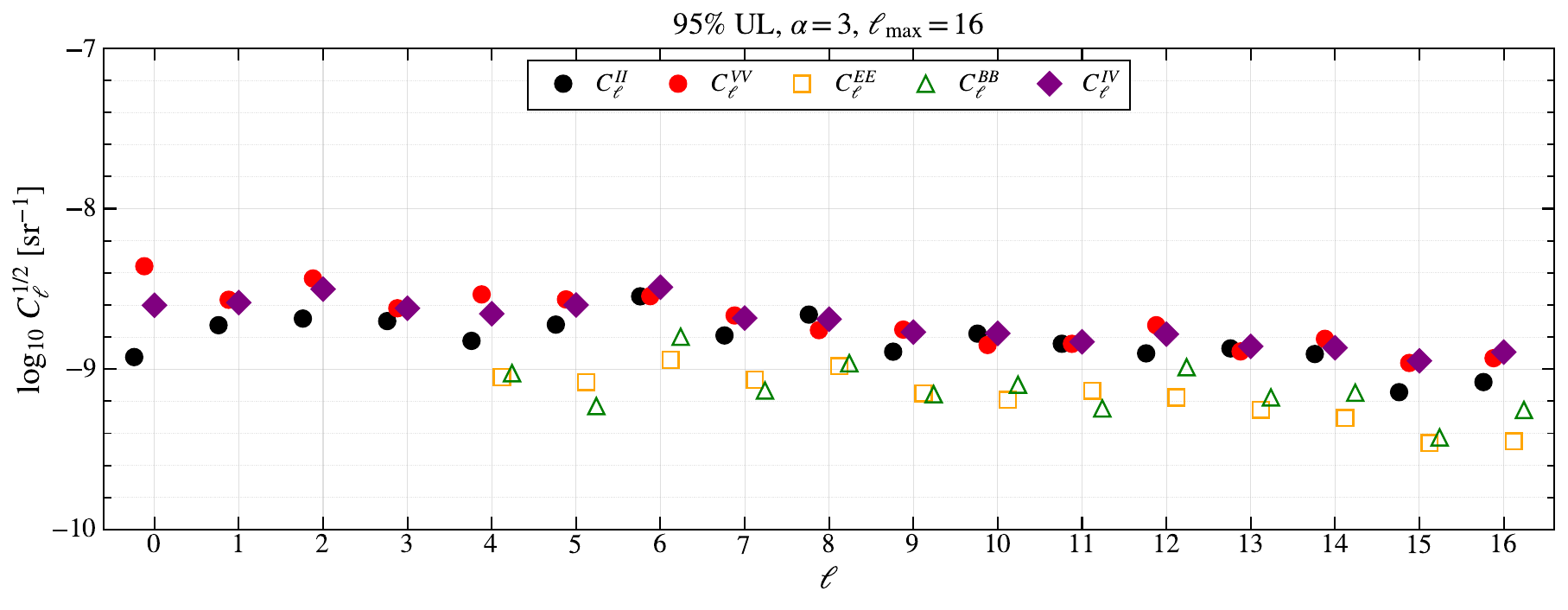}
  \caption{Full-Stokes SPH $95\%$ upper limits
  $\log_{10}C_\ell^{1/2}\,[\mathrm{sr}^{-1}]$ for $\alpha=0$ with
  $\ell_{\max}=3$ on top and $\alpha=3$ with $\ell_{\max}=16$ on the
  bottom, complementing the $\alpha=2/3$ result of
  Fig.~\ref{fig:o3_wide_results}. The $E$ and $B$ spectra start at
  $\ell=4$.}
  \label{fig:o3_sph_ul}
\end{figure*}

\begin{figure*}[htbp]
  \centering
  \includegraphics[width=\linewidth]{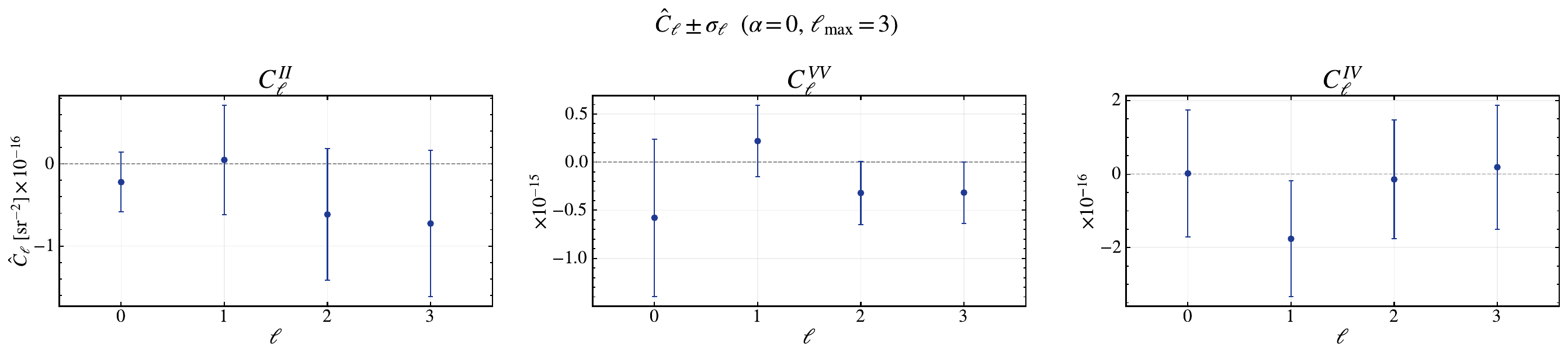}\\[1.5mm]
  \includegraphics[width=\linewidth]{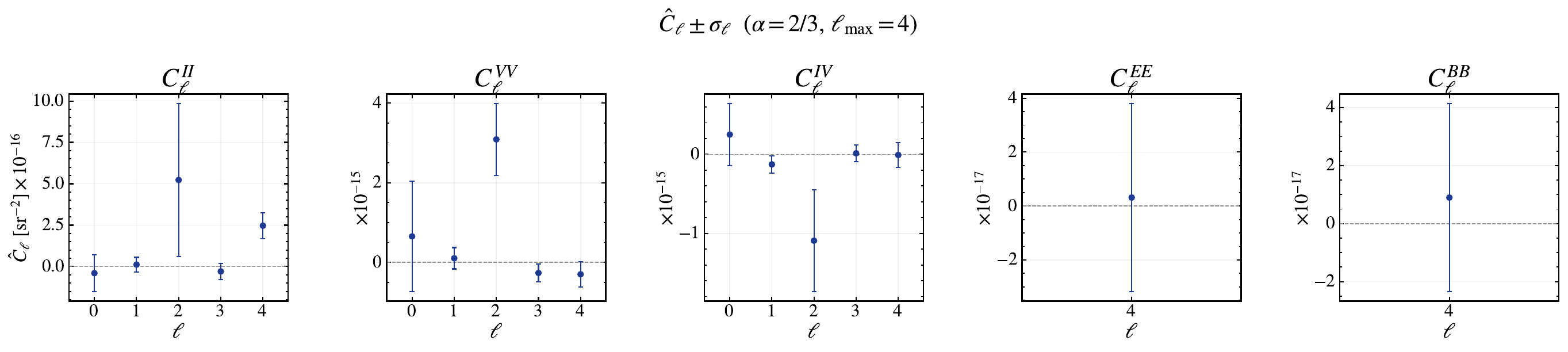}\\[1.5mm]
  \includegraphics[width=\linewidth]{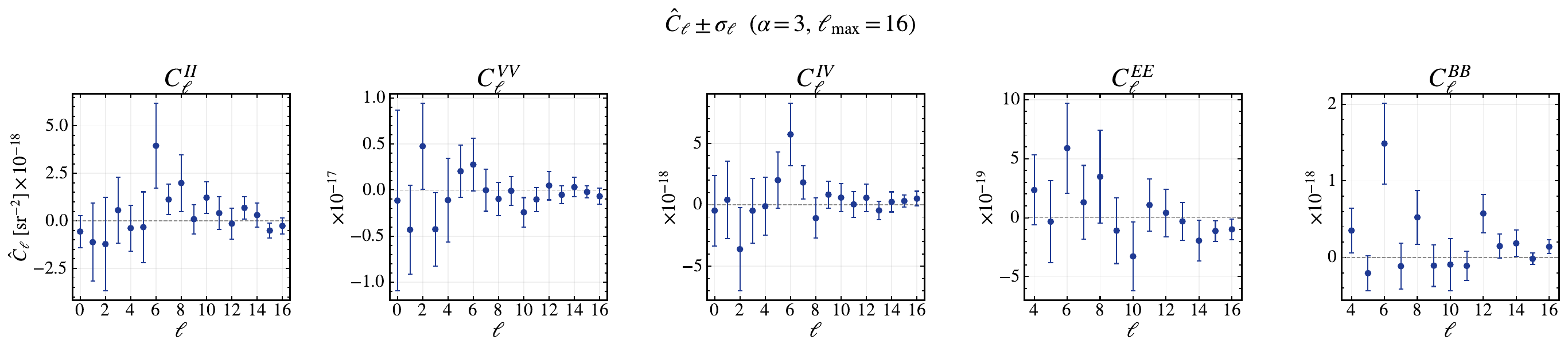}
  \caption{SPH angular-power point estimates
  $\hat C_\ell\pm\sigma_\ell$ on a linear scale, where the estimator
  can be negative; rows are $\alpha=0$, $2/3$ and $3$ from top to
  bottom. All spectra are consistent with zero at the $2\sigma$ level,
  so only upper limits are quoted. The $E$ and $B$ spectra are defined
  for $\ell\ge4$.}
  \label{fig:o3_sph_clest}
\end{figure*}

The angular-power upper limits for the auto spectra use the same flat
non-negative Bayesian construction as the pixel analysis, with the
calibration marginalized in the same way. For the $IV$ cross spectrum,
which is signed, we quote the corresponding $95\%$ bound on
$|C^{IV}_\ell|$. Since $C_\ell\propto\lambda^2$, the per-baseline
calibration factors $\lambda_{\I\J}$ enter the posterior through an
$\ell$-dependent response
\begin{equation}
b_\ell(\lambda)=\frac{\sum_{ab}\big|\sum_{\I\J}\lambda_{\I\J}\,(G^{\I\J}_\ell)_{ab}\big|^{2}}
  {\sum_{ab}\big|\sum_{\I\J}\,(G^{\I\J}_\ell)_{ab}\big|^{2}},
  \qquad G^{\I\J}=\Sigma\,\mathcal F_{\I\J},
\end{equation}
that weights each baseline's contribution to the multipole; the
indices $a,b$ run over the entries of the $(2\ell+1)$-mode block
$G^{\I\J}_\ell$. The posterior is
\begin{equation}
  P(C_\ell\mid\hat C_\ell)\propto\big\langle\mathcal N(\hat C_\ell;\,
  b_\ell(\lambda)\,C_\ell,\,\sigma_\ell^{2})\big\rangle_\lambda,
  \qquad C_\ell\ge 0,
\end{equation}
with the per-detector $\lambda_\I$ marginalized by Monte Carlo as
for the pixel limits; its $95\%$ quantile is the upper limit.
Fig.~\ref{fig:o3_sph_ul} gives the full-Stokes angular-power upper limits for $\alpha=0$ and $\alpha=3$, and
Fig.~\ref{fig:o3_sph_clest} the corresponding point estimates.

\black{Fig.~\ref{fig:lvk_vs_full} compares the analysis of \cite{KAGRA:2021mth} with the full-Stokes search on the same folded O3 data \cite{LVK:O3FoldedData}. In the pixel basis, the upper limits are only $\sim 10\%$ larger than the \cite{KAGRA:2021mth} results since the per pixel inversion of the $4\times 4$ Fisher matrix is well conditioned. In the spherical harmonics basis, the upper limits on $C^{1/2}_\ell$ are $\sim 3$ times larger than the \cite{KAGRA:2021mth} results reflecting the larger variance of the polarized search. SNR maps of the two analyses show the same structure. 
We also compare our results for isotropic parity violation with
\cite{Martinovic:2021hzy}, which reports $\Omega^{95\%}_{\rm ref}=
7 \times 10^{-9}$ for a constant polarization degree and $4.9
\times 10^{-9}$ when the polarization degree is marginalized over
spectral indices. We convert our $C^{VV}_0$ at $\alpha=2/3$ to
$\Omega^{V}= \sqrt{4 \pi C^{VV}_0} \simeq 1.6 \times 10^{-8}$.
That search bounds the intensity amplitude of a partially
polarized background while ours bounds the circular-polarization
amplitude itself, and the two agree at the order-of-magnitude
level at which they are comparable.}

\FloatBarrier
\section{CBC Population and Shot Noise} \label{sec:population}

\black{The population that was used for the case study in the main body of
this paper is constructed in the following manner. We assume an
observation period of three years ($T_{\rm obs}=3\,\mathrm{yr}$). The
BBH population is drawn from the GWTC-5.0 population
posterior~\cite{LIGOScientific:2026ctl}, with the mass model of
Eqs.~(B10)--(B14) of Ref.~\cite{LIGOScientific:2026ctl} and a
Madau--Dickinson redshift evolution~\cite{Madau:2014bja} with
posterior parameters. The adopted posterior draw has
$R_0^{\mathrm{BBH}} = 18\;\mathrm{Gpc^{-3}yr^{-1}}$ at $z=0$. For the
BNS population we assume the Madau--Dickinson shape}
\begin{equation} 
\psi(z) = \frac{(1+z)^{2.7}}{1+((1+z)/(1+z_p))^{5.6}}, 
\end{equation} with a peak redshift of $z_p=1.9$ and take $R_0^{\mathrm{BNS}} = 100\;\mathrm{Gpc^{-3}yr^{-1}}$ at
\black{$z=0$ which is within the 90$\%$ interval of Ref.~\cite{LIGOScientific:2026ctl}. The BNS component masses are drawn from a Gaussian with a mean
of $1.33\,M_\odot$ and a standard deviation of $0.09\,M_\odot$,
truncated to $[1.1, 2.0]\,M_\odot$. Every event has a randomly
selected sky position, a random inclination $\cos\iota\in[-1,1]$ and
a random polarization angle $\psi\in[0,\pi)$. Each catalog is an
independent Poisson draw of the population with about
$1.8\times10^{5}$ BBH events out to $z=10$, and we use 100 catalogs
throughout.} The inspiral phase of each event is truncated at
$f_{\mathrm{ISCO}} = c^3/(6^{3/2}\pi G M_{z,\mathrm{tot}})$, where
$M_{z,\mathrm{tot}} = (m_1+m_2)(1+z)$. \black{The detector network consists of two 15 km L-shaped Einstein
Telescope interferometers located at Sos Enattos (Sardinia) and the
Meuse-Rhine Euroregion and one 40 km Cosmic Explorer at current LIGO-Hanford configuration. Site
coordinates for the ET and arm azimuths follow S3 and R3 of Ref.~\cite{Ebersold:2024hgp}. We use
the 15 km ET sensitivity of Ref.~\cite{ET_0304B_22} and the
40 km CE sensitivity of Ref.~\cite{CE_T2000017}}. For each event, the
network SNR is calculated as
$\rho_{\mathrm{net}} = \sqrt{\sum_d \rho_d^2}$, where the individual
per-detector SNR includes the antenna-pattern response $F^+_d$,
$F^\times_d$ evaluated at the source sky position and the GPS time of
the segment:

\begin{equation}\label{eq:snr}
    \rho_d^2 = 4\int_{f_{\min}}^{f_{\mathrm{ISCO}}} \frac{\left(F_d^{+2}\,h_+^2 + F_d^{\times 2}\,h_\times^2\right)}{S_n^{(d)}(f)}\,df\,,
\end{equation}
with $h_+^2 \propto \left[(1+\cos^2\iota)/2\right]^2$ and $h_\times^2 \propto \cos^2\iota$ encoding the inclination dependence. \black{Events with $\rho_{\mathrm{net}}$ above a threshold $\rho_{\rm th}$
are treated as individually resolved and removed from the residual
population, and we consider $\rho_{\rm th}=8$, $20$ and $50$. }

The
fractional energy density contributed by each remaining event is
\begin{equation}
    \Omega^I_\alpha(f) = \frac{5\pi^{5/3}}{36\,H_0^2\,T_{\mathrm{obs}}} \frac{(G\mathcal{M}_z)^{5/3}}{c^3\,d_L^2}\,f^{2/3}\,g_I\,,
\end{equation}
where $g_I = (1+\cos^2\iota)^2/4 + \cos^2\iota$ is the inclination
factor for the intensity, and the corresponding factor for the
circular polarization is $g_V = \cos\iota\,(1+\cos^2\iota)$.

The shot-noise angular power spectra are computed from the
spherical-harmonic projection of the residual events,
\begin{equation}
    a^X_{\ell m} = \sum_\alpha \omega^X_\alpha\, Y^*_{\ell m}(\hat\Omega_\alpha)\,,\qquad
    C^{XY,\mathrm{shot}}_\ell = \frac{1}{2\ell+1}\sum_m \mathrm{Re}\!\left[a^X_{\ell m}\,a^{Y*}_{\ell m}\right],
\end{equation}
where the sum on $\alpha$ runs over all residual events,
$\omega^I_\alpha = \Omega^I_\alpha(f_{\mathrm{ref}})$ and
$\omega^V_\alpha = \Omega^V_\alpha(f_{\mathrm{ref}})$ are the
per-event amplitudes at the reference frequency, and
$\hat\Omega_\alpha$ is the sky position of event $\alpha$. For a
uniform sky distribution, the shot noise is white in $\ell$ with
$C^{XY,\mathrm{shot}}_\ell = \sum_\alpha \omega^X_\alpha
\omega^Y_\alpha / (4\pi)$.

\begin{figure*}[htbp]
    \includegraphics[width=\linewidth]{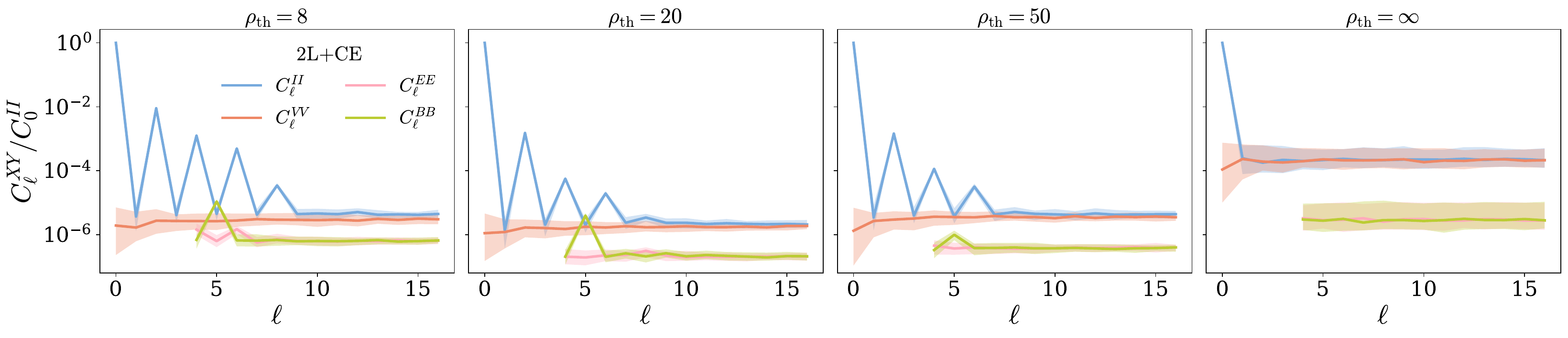}
    \includegraphics[width=\linewidth]{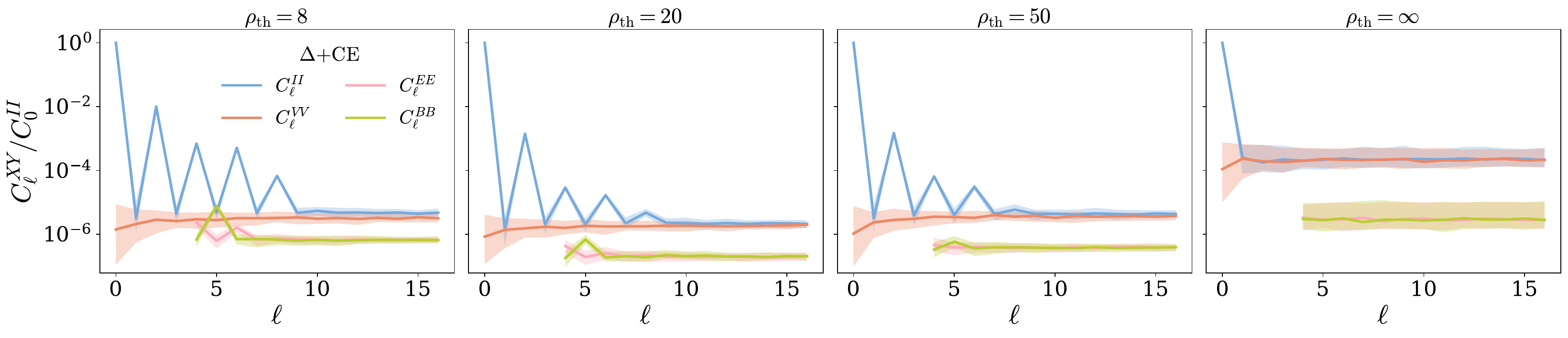}
    \caption{\black{Fractional Angular power spectra of the residual foreground
    at 25 Hz for the CE+ET(2L) (top) and CE+ET($\Delta$) (bottom)
    networks. Lines show the median over 100 population
    realizations with BNS and BBH populations and bands the
    16-84$\%$ range. Panels correspond to
    $\rho_{\rm th}=\{8,20,50\}$ and no removal
    $\rho_{\rm th}=\infty$.}}
    \label{fig:clbands_pop}
\end{figure*}

\black{Fig.~\ref{fig:clbands_pop} shows the residual population
angular power spectra for different polarizations as a function of
different detection thresholds. When no events are assumed to be
detectable the spectra of all polarizations remain flat and
dominated by a few loud sources. We note that since the network is
sensitive to certain directions through Eq.~\eqref{eq:snr} via
$F^{+,\times}_d$, the remaining background has a specific
anisotropic pattern seen in the first few even modes of
$C^{II}_\ell$, which have larger amplitudes. This apparent
anisotropy does not propagate to large $\ell$ values since the
functions $F^{+,\times}_d$ are smooth. This effect also does not
depend much on the underlying detector network and is rather a
general property of the quadrupole shape of the response functions
$F^{+,\times}_d$. Furthermore, the linear polarization spectra
$C^{EE}_\ell$ and $C^{BB}_\ell$ become comparable to the circular
polarization in the residual background.}

For the total CBC background, the inclination averages $\langle g_I
\rangle = 4/5$ and $\langle g_V \rangle = 0$ imply that the
ensemble-mean intensity-circular cross term $\langle
g_I\,g_V\rangle$ vanishes, since $g_I$ is even and $g_V$ is odd in
$\cos\iota$. Thus the ensemble-mean shot-noise bias in
$C^{IV}_\ell$ is zero for an isotropically oriented population,
although finite realizations still contribute variance.
Meanwhile, $C^{VV,\mathrm{shot}}_\ell$ remains nonzero because the
shot-noise power in each Stokes channel scales as $\sum_\alpha
(\omega^X_\alpha)^2$. Since both $C^{II,\mathrm{shot}}_\ell$ and
$C^{VV,\mathrm{shot}}_\ell$ scale in the same way with observing
time, their ratio is independent of observing time and is given by
\begin{align}
\frac{C^{VV,\mathrm{shot}}_\ell}{C^{II,\mathrm{shot}}_\ell}
=
\frac{\langle g_V^2\rangle}{\langle g_I^2\rangle}
\approx 0.97 .
\end{align}
This near-unity ratio implies that polarized shot noise is not
parametrically suppressed (for $\ell \geq 6$) relative to intensity
shot noise, making polarization leakage a relevant systematic for
intensity-only analyses.

\begin{figure}[htbp]
    \centering
    \includegraphics[width=0.5\linewidth]{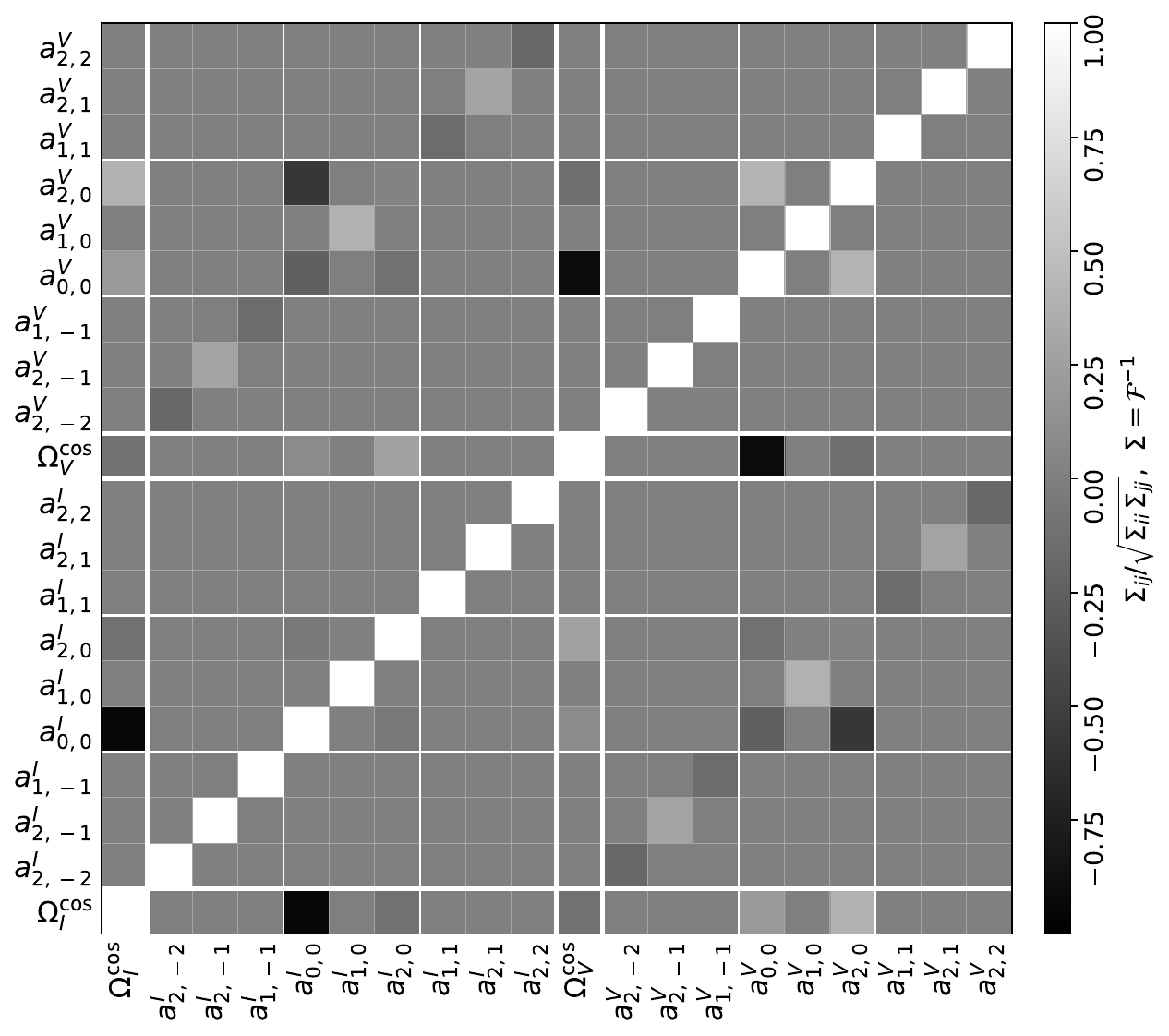}
    \caption{\black{Normalized Fisher matrix of the case study that
    searches for an anisotropic CBC background with spectral index
    $\alpha=2/3$ ($a^{I,V}_{\ell m}$) and an isotropic circularly
    polarized cosmological background ($\Omega^{\rm cos}_I$,
    $\Omega^{\rm cos}_V$) with $\alpha=0$.}}
    \label{fig:fisher_corr}
\end{figure}

\black{We then simulate 100 realizations of the total background observed
by this network over three years. Each realization combines the
resolved and unresolved CBC components of one catalog with an
isotropic cosmological parity violating component with $\alpha=0$,
$\Omega^{\rm cos}_I=10^{-11}$ and $\Omega^{\rm cos}_V=10^{-12}$,
\begin{equation}
    \Omega_X(f,\nhat) = \Omega^{\rm CBC,res}_X(f,\nhat)
    + \Omega^{\rm CBC,unres}_X(f,\nhat)
    + \Omega^{\rm cos}_X(f)\,,
    \label{eq:cs_decomp}
\end{equation}
for each Stokes parameter $X$. The resolved events are assumed to
be perfectly known. This is the case study of the main text,
whose recovered spectra are shown in Fig.~\ref{fig:3g_forecast}.
Each realization is analyzed with the full-Stokes model
\begin{equation}\label{eq:cs_model}
    \Omega^{\rm model}_X(f,\nhat)
    = \left(\frac{f}{f_{\rm ref}}\right)^{2/3}
    \sum_{\ell\leq2,\,m} a^X_{\ell m}\, Y_{\ell m}(\nhat)
    + \Omega^{\rm cos}_X\,,
    \qquad X \in \{I, V\},
\end{equation}
where the multipoles $a^X_{\ell m}$ with $\ell\leq2$ carry the CBC
spectral shape $\alpha=2/3$ and the cosmological amplitudes
$\Omega^{\rm cos}_I$ and $\Omega^{\rm cos}_V$ are isotropic with a
flat spectrum and are quoted as constant angular densities, so
that $C^{XX}_0 = 4\pi(\Omega^{\rm cos}_X)^2$, and with two
restricted models that mirror the current searches, an intensity-only fit of the same multipoles for
the angular spectra and a monopole-only fit of the CBC monopole and
the two cosmological amplitudes for the parity search, which we
refer to as the standard monopole search below. The correlation
matrix of the full-Stokes parameter estimates is shown in
Fig.~\ref{fig:fisher_corr}.}

\black{The even-$\ell$ selection effect matters for the case study because from Fig.~\ref{fig:fisher_corr} we observe a non-zero coupling between $\Omega^{\rm cos}_V$ and the $\ell=2$ modes. Therefore, even though the coupling between $\Omega^{\rm cos}_V$ and $a^I_{00}$ is stronger, which also reflects the fact that multiple isotropic components with different spectral indices can create a bias if not jointly modeled, the anisotropy of the residual background could be a more relevant systematic for parity
violating searches of cosmological signals in the next generation detectors than the direct astrophysical shot noise in $a^{V}_{00}$.}

\begin{figure*}[htbp]
    \centering
    \includegraphics[width=\linewidth]{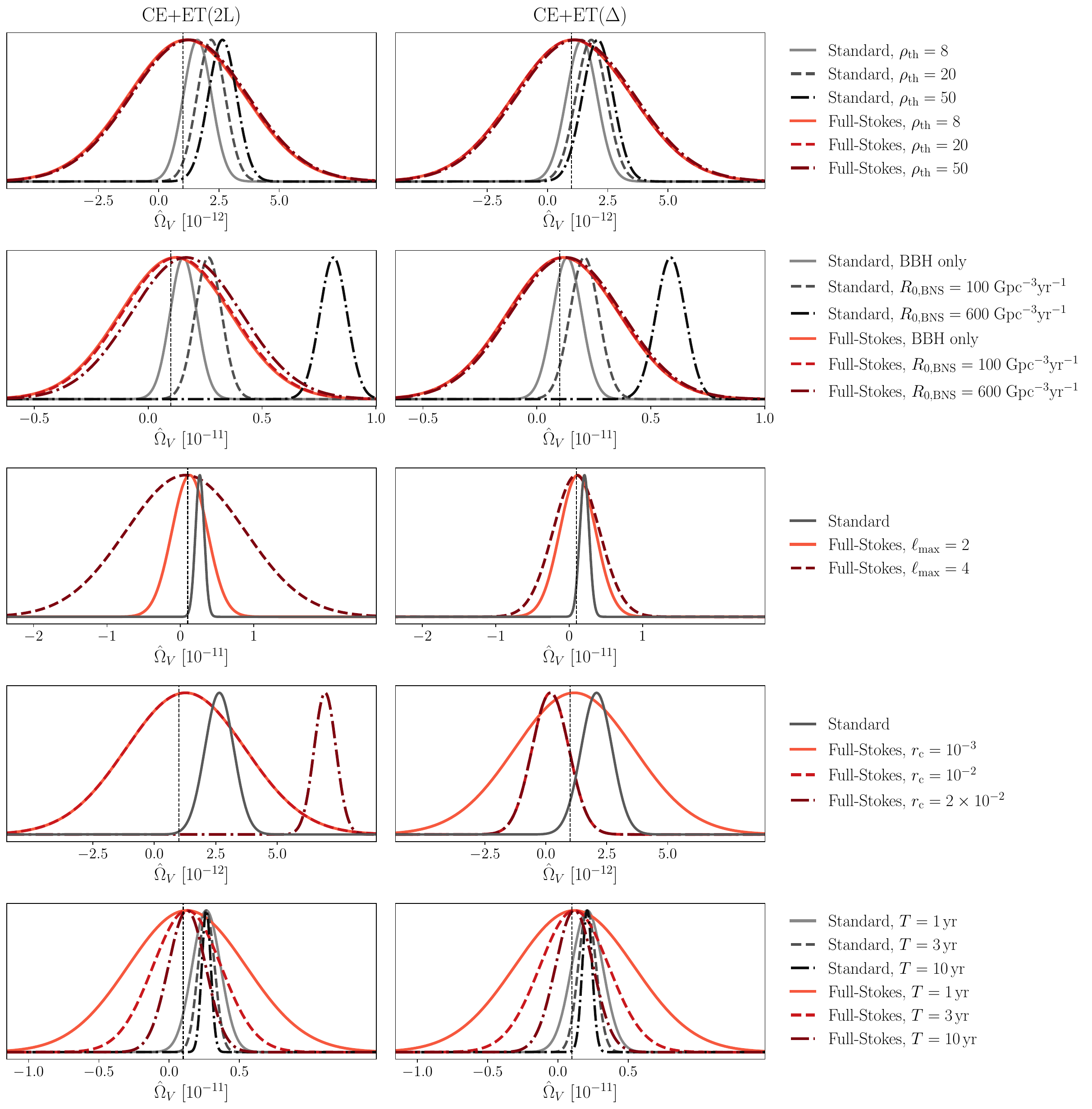}
    \caption{\black{Distribution of the recovered cosmic $V$
    amplitude for the standard (grey) and the full-Stokes (red)
    searches with CE+ET(2L) (top) and CE+ET($\Delta$) (bottom)
    networks. Panels vary the individual CBC detectability
    threshold, the populations, $\ell_{\rm max}$, the
    regularization cutoff and the observation time. The dashed line
    marks the simulated value.}}
    \label{fig:onepanel_robust}
\end{figure*}

\black{In Fig.~\ref{fig:onepanel_robust} we show the distributions of the
recovered cosmic $V$ amplitude over the 100 population realizations
and the detector noise, for the standard monopole search and for
the full-Stokes fit, for both networks. The panels vary one
ingredient of the case study of the main text at a time, the
detection threshold, the population, the analysis $\ell_{\max}$,
the regularization cutoff and the observation time. Apart from the
over-regularized case discussed below, the full-Stokes fit recovers
the injected amplitude in all panels with a bias below a third of
its uncertainty. The standard search is biased in every variant,
and the size of the bias tracks the residual foreground. In the
case-study configuration it is comparable to the injected
amplitude, it grows to several times the injected amplitude when
the BNS rate is raised to $R_0=600\;\mathrm{Gpc^{-3}yr^{-1}}$, and
it shrinks when the detection threshold is lowered to $\rho_{\rm
th}=8$. The uncertainty decreases as $1/\sqrt{T_{\rm obs}}$ while
the bias does not, so the significance of the bias grows with
observing time, from about $1.6\sigma$ at 1 year to about $5\sigma$ at 10 years for the CE+ET(2L) network. Enlarging $\ell_{\max}$ widens the full-Stokes distribution without introducing a bias. The recovery is stable under moderate regularization, while too strong a cutoff removes informative modes, shrinks the nominal uncertainty and biases even the
full-Stokes fit, so in practice the cutoff is chosen inside the plateau where the
recovery does not change. The standard search therefore cannot separate a parity violating signal from the foreground, and its bias only becomes more significant with longer observation.}

\section{Polarization leakage bias derivation} \label{sec:bias_derivation}

Here we derive the leakage bias $\Delta C^{\mathrm{pol}}_\ell$ noted in the main text. The $I$-only clean map $\hat a^I_{\ell m} = \sum_{\ell'}(\mathcal{F}_{II})^{-1}_{\ell\ell',m}\,\hat{\mathcal{X}}^I_{\ell'm}$ absorbs polarization power through the dirty map. \black{The derivation parallels the noise bias analysis of \cite{Kouvatsos:2023bgd} and extends it to the bias produced by unmodeled polarized sources covering persistent and transient sources.} After sidereal averaging the Fisher matrix becomes diagonal in azimuthal index $m$ (i.e., $\mathcal{F}_{\ell m,\ell' m'} \propto \delta_{mm'}$), since the Earth's rotation averages out the $m \neq m'$ couplings. Under this $m$-diagonal structure, the signal content of this map is
\begin{equation}
    \langle \hat a^I_{\ell m}\rangle_N = \sum_p\sum_X\sum_{\ell'}\mathcal{L}^{X,(p)}_{\ell\ell',m}\;\Omega^{X,p}_{\ell'm}\,,
\end{equation}
where $p\in\{s,c\}$ labels two classes of sources: the broadband ($s$, ``shot'') population consists of transient CBC events whose signals are short-lived compared to the observation time and span a wide frequency range, while the quasi-monochromatic ($c$, ``continuous'') population consists of persistent narrowband emitters such as rotating neutron stars. The index $X\in\{I,V,+,-\}$ runs over Stokes parameters, and the leakage matrix
\begin{equation}
    \mathcal{L}^{X,(p)}_{\ell\ell',m} = \sum_{\ell''} \bigl(\mathcal{F}_{II}\bigr)^{-1}_{\ell\ell'',m}\,\mathcal{F}^{IX,(p)}_{\ell''\ell',m}
\end{equation}
couples the pipeline's $I$-template to the Stokes-$X$ content of population~$p$. For the broadband population, $\mathcal{F}^{IX,(s)}$ takes the form of Eq.~\eqref{eq:fish1} summed over the full frequency band. For the quasi-monochromatic population, the broadband spectral weight is replaced by a sum of narrow frequency bins centered on each source frequency. Defining the $\Omega$-normalized response
\begin{equation}
    \tilde\gamma^X_{\ell m}(f)
    \equiv
    \frac{3H_0^2}{4\pi^2 f^3}\,H(f)\,\gamma^X_{\ell m}(f),
\end{equation}
we write
\begin{equation}
    \mathcal{F}^{IX,(c)}_{\ell\ell',m}
    =
    \sum_\alpha w_\alpha\,
    [\tilde\gamma^{I}_{\ell m}(f_\alpha)]^{*}\,
    \tilde\gamma^{X}_{\ell'm}(f_\alpha)\,,
\end{equation}
so that each emitter samples the energy-density response at its own frequency. Because the broadband pipeline inverts the full-band $\mathcal{F}_{II}$, which averages the ORF structure over many frequencies, the leakage $\mathcal{L}^{X,(c)}$ probes a fundamentally different frequency weighting and produces a distinct $\ell$-mixing pattern from $\mathcal{L}^{X,(s)}$. Moreover, since the leakage involves the inverse of $\mathcal{F}_{II}$, the mixing grows with $\ell$ as the Fisher matrix becomes increasingly ill-conditioned.

Each Stokes component in a given time segment $t$ decomposes as
\begin{equation}
    \Omega^{X}_{\ell m}(t) = \bar\Omega_s\,\mathcal{A}^{X,s}_{\ell m} + \bar\Omega_c\,\mathcal{A}^{X,c}_{\ell m} + \delta\Omega^{X,\mathrm{shot}}_{\ell m}(t)\,,
\end{equation}
where $\bar\Omega_s$ and $\bar\Omega_c$ are the intensity monopoles of the two populations, $\mathcal{A}^{X,p}_{\ell m}$ are dimensionless fractional anisotropy fields fixed across segments (encoding large scale structure clustering for the broadband population and the persistent sky pattern for continuous emitters), and $\delta\Omega^{X,\mathrm{shot}}_{\ell m}(t)$ is the zero-mean shot-noise fluctuation from the finite number of transient events per segment ($\delta\Omega^{X,\mathrm{shot}}\equiv 0$ for continuous sources). Their correlators are
\begin{align}
    \langle \mathcal{A}^{X,p}_{\ell m}\,\mathcal{A}^{Y,q*}_{\ell'm'}\rangle_S &= \delta_{\ell\ell'}\delta_{mm'}\,\mathcal{C}^{XY,pq}_\ell\,,\label{eq:corr_pers_sm}\\
    \langle \delta\Omega^{X,\mathrm{shot}}_{\ell m}(t)\,\delta\Omega^{Y,\mathrm{shot}*}_{\ell'm'}(t')\rangle_S &= \delta_{tt'}\delta_{\ell\ell'}\delta_{mm'}\,W^{XY}_\tau\,,\label{eq:corr_shot_sm}
\end{align}
where $\mathcal{C}^{XY,pq}_\ell$ is the dimensionless fractional cross-power spectrum between Stokes $X$ of population $p$ and Stokes $Y$ of population $q$, and $W^{XY}_\tau$ is the per-segment shot-noise cross-power in units of $\Omega^2$. The $\delta_{tt'}$ reflects the statistical independence of shot noise across segments. Working in the spherical-harmonic basis and specializing the leakage matrix of Eq.~\eqref{eq:leakage} to separate broadband ($s$) and continuous ($c$) populations, we form $\hat C^{II}_\ell = (2\ell{+}1)^{-1}\sum_m |\hat a^I_{\ell m}|^2$. Substituting the above correlators, the polarization bias evaluates to
\begin{equation}\label{eq:pol_bias}
    \Delta C^{\mathrm{pol}}_\ell = \frac{1}{2\ell{+}1}\sum_{m,\ell'}\sum_{\substack{X,Y \\ \in\{V,+,-\}}}\bigg[\mathcal{L}^{X,(s)}_{\ell\ell',m}\mathcal{L}^{Y,(s)*}_{\ell\ell',m}\!\left(\bar\Omega_s^2\mathcal{C}^{XY,ss}_{\ell'}+\frac{W^{XY}_\tau}{N_{\mathrm{seg}}}\right) + \bar\Omega_c^2\,\mathcal{L}^{X,(c)}_{\ell\ell',m}\mathcal{L}^{Y,(c)*}_{\ell\ell',m}\mathcal{C}^{XY,cc}_{\ell'} + 2\bar\Omega_s\bar\Omega_c\,\mathrm{Re}\!\left[\mathcal{L}^{X,(s)}_{\ell\ell',m}\mathcal{L}^{Y,(c)*}_{\ell\ell',m}\mathcal{C}^{XY,sc}_{\ell'}\right]\bigg],
\end{equation}
where $N_{\mathrm{seg}} = T_{\rm obs}/\tau$, and $\bar\Omega_s$, $\bar\Omega_c$ are the intensity monopoles of the broadband and quasi-monochromatic populations, respectively.

For a single broadband population ($c=0$), Eq.~\eqref{eq:pol_bias} reduces to
\begin{equation}
    \Delta C^{\mathrm{pol}}_\ell = \frac{1}{2\ell{+}1}\sum_{m,\ell'} \sum_{X,Y} \mathcal{L}^{X}_{\ell\ell',m}\,\mathcal{L}^{Y*}_{\ell\ell',m}\;M^{XY}_{\ell'}\,,
\end{equation}
where $M^{XY}_{\ell'} \equiv \bar\Omega_s^2\,\mathcal{C}^{XY,ss}_{\ell'} + W^{XY}_\tau/N_{\mathrm{seg}}$. Both contributions to $M^{XY}$ are positive semi-definite matrices in Stokes space: $\mathcal{C}^{XY,ss}_\ell$ is a covariance matrix by construction, and $W^{XY}_\tau = \sum_\alpha \omega^X_\alpha\,\omega^{Y*}_\alpha$ is a Gram matrix. Their sum is therefore positive semi-definite. For each $(m,\ell')$, the summand takes the form $\vec{v}^\dagger M\,\vec{v}$ with $v_X = \mathcal{L}^X_{\ell\ell',m}$, which is non-negative. Summing over all $(m,\ell')$ preserves non-negativity, so $\Delta C^{\mathrm{pol}}_\ell \geq 0$. The inequality is strict only if at least one leakage vector has nonzero projection onto the support of $M^{XY}_{\ell'}$.

\end{document}